%% file: vf.tex
\input cwebmac
% vf: a CWEB document by Vincenzo De Florio
% vf 2.0, 28 October 1997.
% Copyright(c) 1996, 1997 Katholieke Universiteit Leuven/ESAT/ACCA
%\magnify{1200}
%\hsize6truein
%\vsize9truein
%\nocon % omit table of contents
%\datethis % print date on listing
\font\Large=cmr12
%\font\large=cmr11
\def\nmr{\hbox{$N\!$M$\!$R}}
\def\ref#1{\hbox{(see [#1])}}
\def\LinkCBxt{{\bf LinkCB\_t}}
\def\VotingFarmxt{{\bf VotingFarm\_t}}
\def\GlobIdxt{{\bf GlobId\_t}}
\def\Optionxt{{\bf Option\_t}}
\def\VFxmsgxt{{\bf VF\_msg\_t}}
\def\Epsilon{\hbox{$\epsilon$}}
\input psfig.tex
%%
%% title
%%

\centerline{ }
\vskip2cm

\centerline{\Large The Voting Farm}
\centerline{\Large A Distributed Class for Software Voting}
\vskip0.5cm
\centerline{Vincenzo De Florio}
\centerline{Universiteit Antwerpen}
\centerline{Department of Mathematics and Computer Science}
\centerline{MOSAIC research group}
\centerline{Middelheimlaan 1, B-2020 Antwerpen}
\vskip0.8cm
%\centerline{\psfig{file=sedes22.eps,height=3cm}}
\centerline{\sc (version revised on April 29, 2015)}
%\centerline{Katholieke Universiteit Leuven,}
%\centerline{Departement Elektrotechniek,}
%\centerline{Afdeling ESAT/ACCA)}
%(version 1.4)
%\centerline{June 7, 1997}
%\centerline{(revised on April 29, 2015.)}

\vfill\eject

%%%%%%

\N{1}{1}VotingFarmTool.

This document describes a class of C functions implementing a distributed
software voting mechanism for EPX [{\sc 11}, {\sc 12}] or similar message passing
multi-threaded environments. Such a tool may be used for example, to set up
a restoring organ [{\sc 9}] i.e., an \nmr{} (i.e., $N$-module
redundant) system with $N$ voters.

In order to describe the tool we start defining its basic building block,
the {\it voter}.

A voter is defined as a software module connected to one
user module and to a farm of fellow voters arranged into
a cliqu\'e.

\vskip 5pt

\centerline{\psfig{file=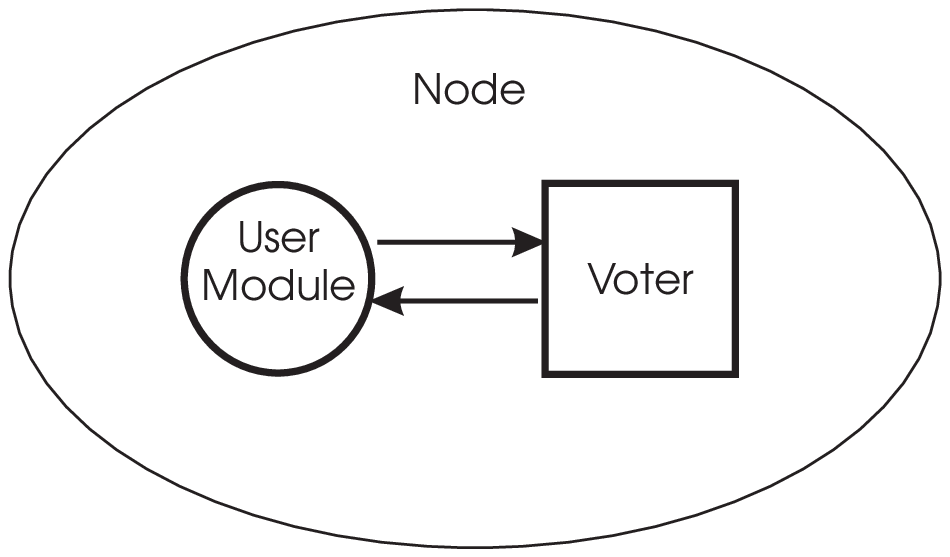}}
\centerline{{\bf Figure 1.} \ninerm A user module and its local voter.}

\vskip 5pt

By means of the functions in the class
the user module is able:

\item{$\bullet$} to create a static ``picture'' of the voting farm, needed
for the set up of the cliqu\'e;
\item{$\bullet$} to instantiate the local voter;
\item{$\bullet$} to send input or control messages to that voter.

\vskip 10pt

\centerline{\psfig{file=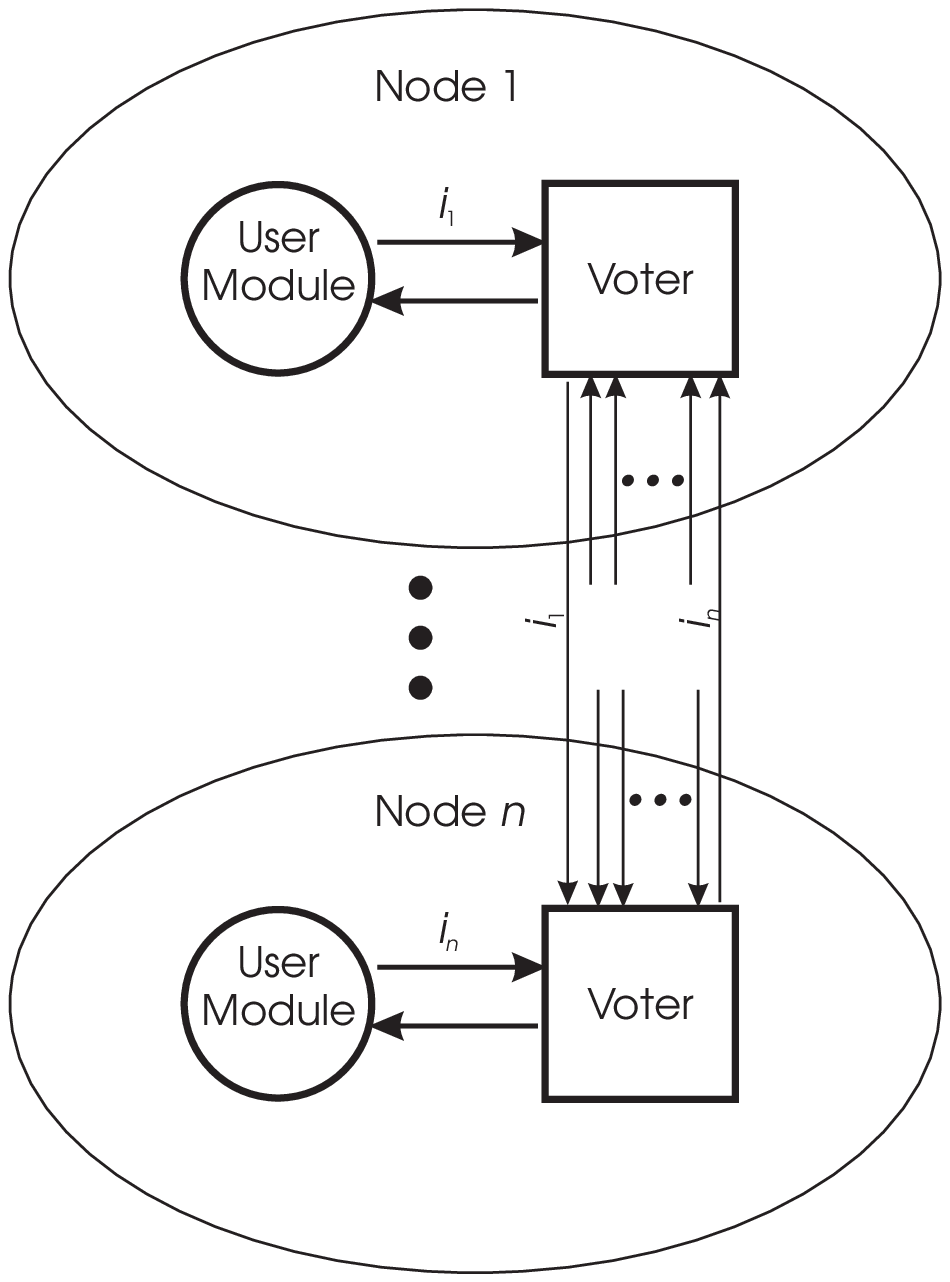}}
\centerline{{\bf Figure 2.}
\ninerm The architecture of the voting farm: each user module connects to
one voter and}
\vskip -3pt
\centerline{\ninerm interacts only with it.
In particular, the user module sends its local
voter only one input value;}
\vskip -3pt
\centerline{\ninerm the voter then broadcasts it across the farm;
then it receives $N-1$ messages from its fellows}
\vskip -3pt
\centerline{\ninerm so to be able to perform the voting.}

\vskip 5pt

No interlocutor is needed other than the local voter.
The other user modules are supposed to create coherent
pictures and instances of voters on other nodes of the machine
and to manage consistently the task of their local intermediary.
All technicalities concerning the set up of the cliqu\'e and the
exchange of messages between the voters are completely transparent
to the user module.
More information about the voting farm may be found in~[{\sc 6}, {\sc
7},
{\sc 8}].

In the following the basic functionalities of the VotingFarm class will be
discussed, namely how to set up a ``passive farm'', or a non-alive (in the
sense
of~[{\sc 4}, {\sc 5}]) topological representation of a
yet-to-be-activated voting farm;
how to initiate the voting farm; how to control the
farm.

\Y\B\X3:Global Variables and \PB{$\#$ \&{include}}'s\X\7
\X4:Voting Farm Declaration\X\6
\X6:Voting Farm Definition\X\6
\X7:Voting Farm Description\X\6
\X11:Voting Farm Activation\X\6
\X14:Voting Farm Control\X\6
\X28:Voting Farm Read\X\6
\X27:Voting Farm Destruction\X\6
\X51:Voting Farm Error Function\X\7
\X29:Voting Algorithms\X\6
\X30:The Voter Function\X\par
\fi

\M{2}Prologue: headers, global variables, etc.
\Y\B\4\D$\.{VF\_MAX\_NTS}$ \5
\T{16}\C{ size of stacks $\equiv$ max value for \PB{\|N} }\par
\B\4\D$\.{VOTING\_FARMS\_MAX}$ \5
\T{64}\C{ max number of simultaneous active voting farms }\par
\B\4\D$\.{VF\_MAX\_INPUT\_MSG}$ \5
\T{512}\C{ max size of an input message }\par
\B\4\D$\.{VF\_MAX\_MSGS}$ \5
\T{10}\C{ max size of the message buffer }\par
\B\4\D$\.{VF\_EVENT\_TIMEOUT}$ \5
\T{10}\C{ \PB{\\{Select}} time-out is 10 seconds }\par
\B\4\D$\.{NO}$ \5
\T{0}\par
\B\4\D$\.{YES}$ \5
\T{1}\Y\par
\B\4\D$\.{E\_VF\_OVERFLOW}$ \5
${-}{}$\T{1}\C{ error conditions }\par
\B\4\D$\.{E\_VF\_CANT\_ALLOC}$ \5
${-}{}$\T{2}\par
\B\4\D$\.{E\_VF\_UNDEFINED\_VF}$ \5
${-}{}$\T{3}\par
\B\4\D$\.{E\_VF\_WRONG\_NODE}$ \5
${-}{}$\T{4}\par
\B\4\D$\.{E\_VF\_GETGLOBID}$ \5
${-}{}$\T{5}\par
\B\4\D$\.{E\_VF\_CANT\_SPAWN}$ \5
${-}{}$\T{6}\C{ \PB{\\{CreateThread}} error }\par
\B\4\D$\.{E\_VF\_CANT\_CONNECT}$ \5
${-}{}$\T{7}\C{ \PB{\\{ConnectLink}} error }\par
\B\4\D$\.{E\_VF\_RECVLINK}$ \5
${-}{}$\T{8}\C{ \PB{\\{RecvLink}} error }\par
\B\4\D$\.{E\_VF\_BROADCAST}$ \5
${-}{}$\T{9}\C{ Invalid input message - can't broadcast }\par
\B\4\D$\.{E\_VF\_DELIVER}$ \5
${-}{}$\T{10}\C{ Invalid output \PB{$\LinkCBxt$} - can't deliver }\par
\B\4\D$\.{E\_VF\_BUSY\_SLOT}$ \5
${-}{}$\T{11}\C{ Duplicated input message }\par
\B\4\D$\.{E\_VF\_WRONG\_VFID}$ \5
${-}{}$\T{12}\par
\B\4\D$\.{E\_VF\_WRONG\_DISTANCE}$ \5
${-}{}$\T{13}\par
\B\4\D$\.{E\_VF\_INVALID\_VF}$ \5
${-}{}$\T{14}\par
\B\4\D$\.{E\_VF\_NO\_LVOTER}$ \5
${-}{}$\T{15}\C{ exactly one voter is mandatorily needed }\par
\B\4\D$\.{E\_VF\_TOO\_MANY\_LVOTERS}$ \5
${-}{}$\T{16}\C{ exactly one voter is mandatorily needed }\par
\B\4\D$\.{E\_VF\_WRONG\_MSG\_NB}$ \5
${-}{}$\T{17}\C{ wrong number of messages }\par
\B\4\D$\.{E\_VF\_SENDLINK}$ \5
${-}{}$\T{18}\C{ \PB{\\{SendLink}} error }\par
\B\4\D$\.{E\_VF\_INPUT\_SIZE}$ \5
${-}{}$\T{19}\C{ inconsistency in the size of the input }\par
\B\4\D$\.{E\_VF\_UNDESCRIBED}$ \5
${-}{}$\T{20}\C{ undescribed \PB{\\{vf}} object }\par
\B\4\D$\.{E\_VF\_INACTIVE}$ \5
${-}{}$\T{21}\C{ inactive \PB{\\{vf}} object }\par
\B\4\D$\.{E\_VF\_UNKNOWN\_SENDER}$ \5
${-}{}$\T{22}\C{ inconsistency---sender unknown }\par
\B\4\D$\.{E\_VF\_EVENT\_TIMEOUT}$ \5
${-}{}$\T{23}\C{ a \PB{\\{Select}} reached time-out }\par
\B\4\D$\.{E\_VF\_SELECT}$ \5
${-}{}$\T{24}\C{ a \PB{\\{Select}} returned an index out of range }\par
\B\4\D$\.{E\_VF\_WRONG\_ALGID}$ \5
${-}{}$\T{25}\C{ AlgorithmID out of range }\par
\B\4\D$\.{E\_VF\_NULLPTR}$ \5
${-}{}$\T{26}\C{ A pointer parameter held \PB{$\NULL$} }\par
\B\4\D$\.{E\_VF\_TOO\_MANY}$ \5
${-}{}$\T{27}\C{ Too many opened voting farms }\Y\par
\B\4\D$\.{VF\_ERROR\_NB}$ \5
\T{28}\C{ number of errors, $\underline{\hbox{plus one}}$ }\par
\B\4\D$\.{VF\_MAX\_FARMS}$ \5
\T{64}\C{ maximum number of farms available }\C{ voting algorithms }\par
\B\4\D$\.{VFA\_EXACT\_CONCENSUS}$ \5
\T{0}\par
\B\4\D$\.{VFA\_MAJORITY}$ \5
\T{1}\par
\B\4\D$\.{VFA\_MEDIAN}$ \5
\T{2}\par
\B\4\D$\.{VFA\_PLURALITY}$ \5
\T{3}\par
\B\4\D$\.{VFA\_WEIGHTED\_AVG}$ \5
\T{4}\par
\B\4\D$\.{VFA\_SIMPLE\_MAJORITY}$ \5
\T{5}\par
\B\4\D$\.{VFA\_SIMPLE\_AVERAGE}$ \5
\T{6}\par
\B\4\D$\.{VF\_SUCCESS}$ \5
\T{1}\par
\B\4\D$\.{VF\_FAILURE}$ \5
\T{0}\Y\par
\B\4\D$\.{VF\_NB\_ALGS}$ \5
\T{7}\C{ nb of algorithms  $\underline{\hbox{plus one}}$ }\C{ default value for
the $\epsilon$ threshold of formalized majority voting }\par
\B\4\D$\.{VFD\_EPSILON}$ \5
\T{5\_-5}\Y\par
\B\4\D$\.{VFP\_INITIALISING}$ \5
\T{0}\par
\B\4\D$\.{VFP\_CONNECTING}$ \5
\T{1}\par
\B\4\D$\.{VFP\_BROADCASTING}$ \5
\T{2}\par
\B\4\D$\.{VFP\_VOTING}$ \5
\T{3}\par
\B\4\D$\.{VFP\_WAITING}$ \5
\T{4}\par
\B\4\D$\.{VFP\_FAILED}$ \5
\T{5}\par
\B\4\D$\.{VFP\_QUITTING}$ \5
\T{6}\par
\fi

\M{3}Two global variables have been supplied for the user to quer1y the
error status of the application and the name of the function
which experienced the error. Their definition constitutes the main part
of the following section.

\Y\B\4\X3:Global Variables and \PB{$\#$ \&{include}}'s\X${}\E{}$\6
\8\#\&{include} \.{<stdio.h>}\6
\8\#\&{include} \.{<stdlib.h>}\6
\8\#\&{include} \.{<stdarg.h>}\6
\8\#\&{include} \.{<sys/root.h>}\6
\8\#\&{include} \.{<sys/logerror.h>}\6
\8\#\&{include} \.{<sys/link.h>}\6
\8\#\&{include} \.{<sys/select.h>}\6
\8\#\&{include} \.{<sys/time.h>}\6
\8\#\&{include} \.{<sys/thread.h>}\6
\8\#\&{ifdef} \.{SERVERNET}\6
\8\#\&{include} \.{"server.h"}\6
\8\#\&{endif}\C{ SERVERNET }\6
\&{int} \\{VF\_error};\C{ global variable for storing the error condition }\6
\&{static} \&{double} ${}\Epsilon\K\.{VFD\_EPSILON};{}$\6
\&{static} \&{double} \\{ScalingFactor}${}\K\T{1.0};{}$\6
\&{int} \\{once}${}\K\T{1};{}$\6
\&{static} \&{int} ${}\\{VF\_RequestId}(\&{int},\39\&{int},\39\&{int});{}$\6
\&{typedef} \&{struct} ${}\{{}$\1\6
\&{unsigned} \&{char} ${}{*}\\{item};{}$\6
\&{int} \\{item\_nr};\2\6
${}\}{}$ \&{cluster\_t};\C{   A flag is attached to each object so that the
object can be logically   ``deleted'' from the list simply setting its status
to \PB{\.{NOT\_PRESENT}}.   Once the list is created, all its elements are
labeled as \PB{\.{PRESENT}};   as the execution goes by, elements are
``logically'' removed from the   list changing their status to \PB{\.{NOT%
\_PRESENT}}.   }\6
\&{typedef} \&{unsigned} \&{char} \&{flag};\6
\&{typedef} \&{struct} ${}\{{}$\1\6
\&{void} ${}{*}\\{object};{}$\6
\&{flag} \\{status};\2\6
${}\}{}$ \&{value\_t};\C{ This part has been added in V1.5. It defines a set of
(redefineable)    symbolic constants representing upper limits for
pre-allocated areas    used exclusively in the {\bf static} version of the
tool.  }\6
\8\#\&{ifndef} \.{VF\_STATIC\_MAX\_INP\_MSG}\6
\8\#\&{define} \.{VF\_STATIC\_MAX\_INP\_MSG} \5\T{64}\6
\8\#\&{endif}\6
\8\#\&{ifndef} \.{VF\_STATIC\_MAX\_LINK\_NB}\6
\8\#\&{define} \.{VF\_STATIC\_MAX\_LINK\_NB} \5\T{16}\6
\8\#\&{endif}\6
\8\#\&{ifndef} \.{VF\_STATIC\_MAX\_VOTER\_INPUTS}\6
\8\#\&{define} \.{VF\_STATIC\_MAX\_VOTER\_INPUTS} \5\T{20}\6
\8\#\&{endif}\6
\8\#\&{ifdef} \.{STATIC}\6
\8\#\&{define} \\{AllocationClass}\1\1 \&{static} \6
$\LinkCBxt*\\{st\_links}[\.{VF\_STATIC\_MAX\_LINK\_NB}];{}$\6
${}\Optionxt\\{st\_options}[\.{VF\_STATIC\_MAX\_LINK\_NB}];{}$\7
\&{double} \\{st\_VFA\_sum};\C{used in VFA algorithms }\6
\&{double} \\{st\_VFA\_weight}[\.{VF\_STATIC\_MAX\_INP\_MSG}];\6
\&{double} ${}\\{st\_VFA\_squaredist}[\.{VF\_STATIC\_MAX\_INP\_MSG}*\.{VF%
\_STATIC\_MAX\_INP\_MSG}];{}$\6
\&{cluster\_t} \\{st\_clusters}[\.{VF\_STATIC\_MAX\_VOTER\_INPUTS}];\6
\&{void} ${}{*}\\{st\_voter\_inputs}[\.{VF\_STATIC\_MAX\_VOTER\_INPUTS}];{}$\6
\&{char} \\{st\_chars}[\.{VF\_STATIC\_MAX\_VOTER\_INPUTS}];\6
\&{value\_t} \\{st\_VFA\_v}[\.{VF\_STATIC\_MAX\_INP\_MSG}];\6
\&{char} \\{st\_voter\_inputs\_data}[\.{VF\_STATIC\_MAX\_VOTER\_INPUTS}][\.{VF%
\_STATIC\_MAX\_INP\_MSG}];\6
\8\#\&{else}\6
\8\#\&{define} \\{AllocationClass}\6
\8\#\&{endif}\6
\&{unsigned} \&{char} \\{st\_VFA\_vote}[\.{VF\_STATIC\_MAX\_INP\_MSG}];\par
\U1.\fi

\N{1}{4}Voting Farm Declaration.
The whole \PB{\\{VotingFarm}} class is built upon type {\bf Voting} {\bf Farm%
\_t},
which plays the same role as the type \PB{\&{FILE}} in the standard class
of C functions for file management: it offers the user a way to refer to some
object from an abstract point of view, masking him/her from all unneeded
information concerning its implementation.
All a user needs to know is that, in order to use a voting farm, he/she has
first to declare an object like follows:
\PB{$\VotingFarmxt*\\{vf};$}

The newly defined \PB{\\{vf}} variable does not describe any valid voting farm
yet;
it is simply a pointer with no object attached to it, exactly the same way
it goes for a \PB{\&{FILE} ${}{*}\\{fp}$} variable which has not been \PB{%
\\{fopen}}'d yet.
For that a special function is supplied: \PB{\\{VF\_open}}, which is discussed
in the next subsection.

Each user module which needs to use a voting farm should declare a
\PB{$\VotingFarmxt$ $*$} variable.

\Y\B\4\X4:Voting Farm Declaration\X${}\E{}$\6
\&{typedef} \&{struct} ${}\{{}$\1\6
\&{int} \\{vf\_id};\6
\&{int} \\{vf\_node\_stack}[\.{VF\_MAX\_NTS}];\6
\&{int} \\{vf\_ident\_stack}[\.{VF\_MAX\_NTS}];\7
${}\LinkCBxt*\\{pipe}[\T{2}];{}$\7
\&{int} \|N;\6
\&{int} \\{user\_thread};\6
\&{int} \\{this\_voter};\6
\&{double} ${}({*}\\{distance}){}$(\&{void} ${}{*},\39{}$\&{void} ${}{*});{}$\6
\&{flag} \\{broadcast\_done};\6
\&{flag} \\{inp\_msg\_got};\6
\&{flag} \\{destroy\_requested};\6
\8\#\&{ifdef} \.{SERVERNET}\7
\\{RTC\_Thread\_t}\\{rtc};\6
\8\#\&{endif}\2\6
${}\}{}$ ${}\&{VotingFarm\_t};{}$\6
\&{static} \&{int} ${}\\{VF\_voter}(\&{VotingFarm\_t}{}$ ${}{*});{}$\6
\&{int} ${}\\{VF\_add}(\&{VotingFarm\_t}{}$ ${}{*},\39\&{int},\39\&{int});{}$\6
\&{void} \\{VF\_perror}(\&{void});\6
\8\#\&{ifdef} \.{STATIC}\6
${}\&{VotingFarm\_t}{}$ \\{Table}[\.{VF\_MAX\_FARMS}];\6
\8\#\&{endif}\7
\&{void} ${}{*}{}$\\{memdup}(\&{void} ${}{*}\|p,\39{}$\&{size\_t} \\{len})\1\1%
\2\2\6
${}\{{}$\1\6
\&{void} ${}{*}\|q\K\\{malloc}(\\{len});{}$\7
\&{if} (\|q)\1\5
${}\\{memcpy}(\|q,\39\|p,\39\\{len});{}$\2\6
\&{return} \|q;\6
\4${}\}{}$\2\par
\U1.\fi

\M{5}In the above, \PB{\\{vf\_id}} is a unique integer which identifies a
voting
context, \PB{\\{vf\_node\_stack}} is a stack of node ids (a voter thread will
be spawned on each of them), \PB{\\{vf\_ident\_stack}} is a stack of thread id,
\PB{\|N} is a stack pointer (only one stack pointer is needed because the
two stacks evolve in parallel), and \PB{\\{user\_thread}} is the thread id of
the caller (or user) module.
\PB{\\{pipe}} is a couple of pointers to \PB{$\LinkCBxt$}, used to communicate
between the user module and the local voter.
\PB{\\{this\_voter}} is the entry of the current voter.

\fi

\N{1}{6}Voting Farm Definition.
A voting farm variable \PB{\\{vf}} can be defined by means of function \PB{%
\\{VF\_open}}
e.g.,  as follows:

\vskip 5pt

\item{}\hskip 35pt{\bf (1)} \PB{\&{VotingFarm\_t} ${}{*}\\{vf};$}
\item{}\hskip 35pt{\bf (2)} \PB{$\\{vf}\K\\{VF\_open}(\T{5},\\{distance});$}

\vskip 5pt

After this statement has been executed, an object has been allocated, some
initializations have occurred, and the address of the newly created object
has been returned into \PB{\\{vf}}. The number the user supplies as the
argument
of \PB{\\{VF\_open}} is an integer which univokely represents
the current voting farm and in fact distinguishes it from all other voting
farms which possibly will be used at the same times---we may call it
a VotingFarm-id. \PB{\\{distance}} is an arbitrary metric i.e., a function
which gets two pointers to opaque objets, computes a ``distance'', and returns
this value as a positive real number.

Each user module which needs to assemble the same voting farm should
actually execute a \PB{\\{VF\_open}} statement with the same number as an
argument---it is the user's responsibility to do like this. Likewise, coherent
behaviour requires that the same metric function is referenced as second
parameter
of \PB{\\{VF\_open}}.

\PB{\\{VF\_open}} returns \PB{$\NULL$} in case of error; otherwise, it returns
a pointer to a valid object.

\Y\B\4\X6:Voting Farm Definition\X${}\E{}$\6
\&{VotingFarm\_t} ${}{*}{}$\\{VF\_open}(\&{int} \\{vf\_id}${},\39{}$\&{double}
${}({*}\\{distance}){}$(\&{void} ${}{*},\39{}$\&{void} ${}{*})){}$\1\1\2\2\6
${}\{{}$\6
\8\#\&{ifdef} \.{STATIC}\1\6
\&{static} \&{int} \\{vf\_max\_farms};\6
\8\#\&{endif}\7
\\{AllocationClass}\1\1\6
\&{VotingFarm\_t} ${}{*}\\{vf};{}$\6
\&{static} \&{char} ${}{*}\.{VFN}\K\.{"VF\_open"};\2\2{}$\6
\&{if} ${}(\\{vf\_id}\Z\T{0}){}$\5
${}\{{}$\1\6
${}\\{LogError}(\.{EC\_ERROR},\39\.{VFN},\39\.{"Illegal\ VotingFarm\ }\)%
\.{Identifier\ (\%d)\ ---\ }\)\.{should\ be\ greater\ th}\)\.{an\ 0."},\39\\{vf%
\_id});{}$\6
${}\\{VF\_error}\K\.{E\_VF\_WRONG\_VFID};{}$\6
\&{return} ${}\NULL;{}$\6
\4${}\}{}$\2\6
\8\#\&{ifndef} \.{STATIC}\7
\&{if} ${}((\\{vf}\K{}$(\&{VotingFarm\_t} ${}{*}){}$ \\{malloc}(\&{sizeof}(%
\&{VotingFarm\_t})))${}\E\NULL){}$\5
${}\{{}$\1\6
${}\\{LogError}(\.{EC\_ERROR},\39\.{VFN},\39\.{"Memory\ Allocation\ E}\)%
\.{rror."});{}$\6
${}\\{VF\_error}\K\.{E\_VF\_CANT\_ALLOC};{}$\6
\&{return} ${}\NULL;{}$\6
\4${}\}{}$\2\6
\8\#\&{else}\C{ if \PB{\.{STATIC}} is defined, then we fetch the next entry
from array `\PB{\\{Table}}'    }\6
\&{if} ${}(\\{vf\_max\_farms}<\.{VF\_MAX\_FARMS}){}$\1\5
${}\\{vf}\K{\AND}\\{Table}[\\{vf\_max\_farms}\PP];{}$\2\6
\&{else}\5
${}\{{}$\1\6
${}\\{LogError}(\.{EC\_ERROR},\39\.{VFN},\39\.{"Too\ many\ farms."});{}$\6
${}\\{VF\_error}\K\.{E\_VF\_TOO\_MANY};{}$\6
\&{return} ${}\NULL;{}$\6
\4${}\}{}$\2\6
\8\#\&{endif}\6
\&{if} ${}(\\{distance}\E\NULL){}$\5
${}\{{}$\1\6
${}\\{LogError}(\.{EC\_ERROR},\39\.{VFN},\39\.{"Invalid\ Metric\ Func}\)\.{tion%
\ (NULL)."});{}$\6
${}\\{VF\_error}\K\.{E\_VF\_WRONG\_DISTANCE};{}$\6
\&{return} ${}\NULL;{}$\6
\4${}\}{}$\2\7
${}\\{vf}\MG\|N\K\T{0}{}$;\C{ zero the stack pointer }\6
${}\\{vf}\MG\\{vf\_id}\K\\{vf\_id}{}$;\C{ record the vf id }\6
${}\\{vf}\MG\\{distance}\K\\{distance}{}$;\C{ record the function pointer }\6
${}\\{vf}\MG\\{user\_thread}\K\\{vf}\MG\\{this\_voter}\K{-}\T{1};{}$\6
${}\\{LocalLink}(\\{vf}\MG\\{pipe}){}$;\C{ Create a means for communicating
with the local voter }\6
\&{return} \\{vf};\6
\4${}\}{}$\2\par
\U1.\fi

\N{1}{7}Voting Farm Description.
Once a \PB{\&{VotingFarm\_t}} pointer has been created and once an object has
been
correctly defined and attached to that pointer, the user needs to describe
the farm: how many voters are needed, where they should be
placed, how to refer to each voter, and so on. This is accomplished by means
of function \PB{\\{VF\_add}}.  If the voting farm consists of $N$ voters, then
the user shall call \PB{\\{VF\_add}} $N$ times; each call describes a voter
by attaching a couple $(n,t)$ to it, where $n$ is the node of the voter
and $t$ is its thread identifier. As an example, the following statements:

\vskip 5pt

\item{}\hskip 35pt{\bf (1)} \PB{\&{VotingFarm\_t} ${}{*}\\{vf};$}
\item{}\hskip 35pt{\bf (2)} \PB{$\\{vf}\K\\{VF\_open}(\T{5},\\{distance});$}
\item{}\hskip 35pt{\bf (3)} \PB{$\\{VF\_add}(\\{vf},\T{15},\\{tid1});$}
\item{}\hskip 35pt{\bf (4)} \PB{$\\{VF\_add}(\\{vf},\T{21},\\{tid2});$}
\item{}\hskip 35pt{\bf (5)} \PB{$\\{VF\_add}(\\{vf},\T{4},\\{tid5});$}

\vskip 5pt

\noindent
declare (line {\bf (1)}), define (line {\bf (2)}), and
describe (lines {\bf (3)}--{\bf (5)}) voting farm number 5. A triple of voters
has been
proposed; voter 0, identified by the couple $(n,t)=(15,\PB{\\{tid1}})$,
voter 1, or couple $(21,\PB{\\{tid2}})$, and voter 2, or couple $(4,\PB{%
\\{tid5}})$.

Again it is the user's responsibility to operate in a coherent, consistent
way during these phases: in this case, he or she needs to declare the farm
in exactly the same order, with exactly the same cardinality, with the
same attributes on all nodes. Exactly one node-id has to be present and
equal to the number of the current node.

So far no thread has been launched, and no distributed action has
taken place---therefore we talk of ``passive voting farms'' for
farms that have been only declared, defined, and described, but not
activated yet.

\PB{\\{VF\_add}} returns a negative integer in case of error; otherwise, it
returns
zero.

\Y\B\4\X7:Voting Farm Description\X${}\E{}$\6
\&{int} \\{VF\_add}(\&{VotingFarm\_t} ${}{*}\\{vf},\39{}$\&{int} \\{node}${},%
\39{}$\&{int} \\{identifier})\1\1\2\2\6
${}\{{}$\1\6
\&{static} \&{int} \\{this\_node};\C{ set function name }\6
\&{static} \&{char} ${}{*}\.{VFN}\K\.{"VF\_add"};{}$\7
\&{if} ${}(\\{this\_node}\E\T{0}){}$\1\5
${}\\{this\_node}\K\.{GET\_ROOT}(\,)\MG\\{ProcRoot}\MG\\{MyProcID}{}$;\2\7
\X8:Has \PB{\\{vf}} been defined?\X\6
\X9:Is \PB{\\{vf}} a valid object?\X\7
${}\\{vf}\MG\\{vf\_node\_stack}[\\{vf}\MG\|N]\K\\{node};{}$\6
${}\\{vf}\MG\\{vf\_ident\_stack}[\\{vf}\MG\|N]\K\\{identifier};{}$\6
\&{if} ${}(\\{node}\E\\{this\_node}){}$\1\6
\&{if} ${}(\\{vf}\MG\\{this\_voter}<\T{0}){}$\1\5
${}\\{vf}\MG\\{this\_voter}\K\\{vf}\MG\|N{}$;\C{ store the current value of the
stack pointer }\2\6
\&{else}\5
${}\{{}$\1\6
${}\\{LogError}(\.{EC\_ERROR},\39\.{VFN},\39\.{"There\ must\ be\ only\ }\)%
\.{one\ local\ voter."});{}$\6
\&{return} \\{VF\_error}${}\K\.{E\_VF\_TOO\_MANY\_LVOTERS};{}$\6
\4${}\}{}$\2\2\6
${}\\{vf}\MG\|N\PP{}$;\7
\X10:Check stacks growth\X\7
\&{return} \\{VF\_error}${}\K\T{0};{}$\6
\4${}\}{}$\2\par
\U1.\fi

\M{8}Checks if \PB{\\{vf}} holds a valid (non-\PB{$\NULL$}) address.
\Y\B\4\X8:Has \PB{\\{vf}} been defined?\X${}\E{}$\6
\&{if} ${}(\\{vf}\E\NULL){}$\5
${}\{{}$\1\6
${}\\{LogError}(\.{EC\_ERROR},\39\.{VFN},\39\.{"Undefined\ VotingFar}\)\.{m\_t\
Object."});{}$\6
${}\\{LogError}(\.{EC\_ERROR},\39\.{VFN},\39\.{"\\t(A\ VF\_open\ is\ pro}\)%
\.{bably\ needed.)"});{}$\6
\&{return} \\{VF\_error}${}\K\.{E\_VF\_UNDEFINED\_VF};{}$\6
\4${}\}{}$\2\par
\Us7, 11, 24\ETs30.\fi

\M{9}Checks if \PB{\\{vf}} points to valid data.
\Y\B\4\X9:Is \PB{\\{vf}} a valid object?\X${}\E{}$\6
\&{if} ${}(\\{vf}\MG\\{vf\_id}<\T{0}\V\\{vf}\MG\\{vf\_node\_stack}\E\NULL\V%
\\{vf}\MG\\{vf\_ident\_stack}\E\NULL\V\\{vf}\MG\|N<\T{0}){}$\5
${}\{{}$\1\6
${}\\{LogError}(\.{EC\_ERROR},\39\.{VFN},\39\.{"Corrupted\ or\ Invali}\)\.{d\
VotingFarm\_t\ Objec}\)\.{t."});{}$\6
\&{return} \\{VF\_error}${}\K\.{E\_VF\_INVALID\_VF};{}$\6
\4${}\}{}$\2\par
\Us7\ET30.\fi

\M{10}Check if a stack overflow event has occurred.
\Y\B\4\X10:Check stacks growth\X${}\E{}$\6
\&{if} ${}(\\{vf}\MG\|N\G\.{VF\_MAX\_NTS}){}$\5
${}\{{}$\1\6
${}\\{LogError}(\.{EC\_ERROR},\39\.{VFN},\39\.{"Stack\ Overflow."});{}$\6
${}\\{LogError}(\.{EC\_ERROR},\39\.{VFN},\39\.{"\\t(Increase\ the\ val}\)\.{ue\
of\ VF\_MAX\_NTS;\ cu}\)\.{rrent\ value\ is\ \%d.)"},\39\.{VF\_MAX\_NTS});{}$\6
\&{return} \\{VF\_error}${}\K\.{E\_VF\_OVERFLOW};{}$\6
\4${}\}{}$\2\par
\U7.\fi

\N{1}{11}Voting Farm Activation.
After having described a voting farm, next step is turning that passive
description into a ``living'' (active) object: this is accomplished by means
of function \PB{\\{VF\_run}} which simply spawns the local voter and
connects to it. Any inconsistency like e.g., zero or two local voters
are managed at this point and results in specific error messages.
The one argument \PB{\&{VotingFarm\_t} ${}{*}\\{vf}$} is passed to the newly
created thread.

\PB{\\{VF\_run}} returns a negative integer in case of error; otherwise,
it returns zero.

\Y\B\4\X11:Voting Farm Activation\X${}\E{}$\6
\&{int} \\{VF\_run}(\&{VotingFarm\_t} ${}{*}\\{vf}){}$\1\1 $\{$ %
\\{AllocationClass}\1\1\6
\&{int} \\{MyProcID};\2\2\6
${}\\{AllocationClass}\GlobIdxt\\{GlobId};{}$\7
\\{AllocationClass}\1\1\6
\&{int} \\{Error};\6
\8\#\&{ifdef} \.{SERVERNET}\6
\&{extern} ${}\LinkCBxt{*}\\{link2server};{}$\6
\8\#\&{endif}\C{ set function name }\6
\&{static} \&{char} ${}{*}\.{VFN}\K\.{"VF\_run"}{}$;\2\2\7
\X8:Has \PB{\\{vf}} been defined?\X\7
\X12:Has \PB{\\{vf}} been described?\X\7
\&{if} ${}(\\{vf}\MG\\{this\_voter}<\T{0}){}$\5
${}\{{}$\1\6
${}\\{LogError}(\.{EC\_ERROR},\39\.{VFN},\39\.{"No\ voter\ has\ been\ d}\)%
\.{efined\ to\ be\ local."});{}$\6
\&{return} \.{E\_VF\_NO\_LVOTER};\6
\4${}\}{}$\2\6
${}\\{MyProcId}\K\.{GET\_ROOT}(\,)\MG\\{ProcRoot}\MG\\{MyProcID}{}$;\C{ Get the
Global ID structure. }\6
\&{if} ${}(\\{GetGlobId}({\AND}\\{GlobId},\39\NULL)\E{-}\T{1}){}$\5
${}\{{}$\1\6
${}\\{LogError}(\.{EC\_ERROR},\39\.{VFN},\39\.{"Cannot\ get\ the\ glob}\)\.{al\
ID\ of\ the\ thread.}\)\.{"});{}$\6
\&{return} \.{E\_VF\_GETGLOBID};\6
\4${}\}{}$\C{ for the time being, this field is treated as a flag      which
tells whether \PB{\\{VF\_run}} has been executed or not      on voting farm %
\PB{\\{vf}}. Its role will be different when      \PB{\\{FT\_Create\_Thread}}
will be used instead of \PB{\\{CreateThread}}.    }\2\6
${}\\{vf}\MG\\{user\_thread}\K\T{0}{}$;\C{ first create a separate thread for
the voter function }\6
\8\#\&{ifdef} \.{SERVERNET}\6
${}\\{LogError}(\.{EC\_MESS},\39\.{VFN},\39\.{"Creating\ thread\ `VF}\)\.{%
\_voter()'\ via\ RTC\_Cr}\)\.{eateLThread()"});$ $\\{vf}\MG\\{rtc}\K\\{RTC%
\_CreateLThread}$ $(\\{link2server},\39\\{vf}\MG\\{vf\_ident\_stack}[\\{vf}\MG%
\\{this\_voter}],\39\.{DIR\_USER\_TYPE},\39\NULL,\39\T{0},\39(\\{RTC\_ptr\_t})%
\\{VF\_voter},\39\\{vf},\39$ \&{sizeof} ( $\LinkCBxt$ $*$ ) )  ;\6
${}\\{LogError}(\.{EC\_MESS},\39\.{VFN},\39\.{"RTC\_CreateLThread()}\)\.{\ has\
been\ executed."});{}$\6
\8\#\&{else}\6
\&{if} ${}(\\{CreateThread}(\NULL,\39\T{0},\39(\&{int}(*)(\,))\\{VF\_voter},%
\39{\AND}\\{Error},\39\\{vf})\E\NULL){}$\5
${}\{{}$\1\6
${}\\{LogError}(\.{EC\_ERROR},\39\.{VFN},\39\.{"Cannot\ start\ a\ vote}\)\.{r\
thread,\ error\ code}\)\.{\ \%d."},\39\\{Error});{}$\6
\&{return} \\{VF\_error}${}\K\.{E\_VF\_CANT\_SPAWN};{}$\6
\4${}\}{}$\2\6
\8\#\&{endif}\C{ SERVERNET }\7
\&{return} \\{VF\_error}${}\K\T{0};$ $\}{}$\par
\U1.\fi

\M{12}This checks whether the farm has been described by means of
at least one call to \PB{\\{VF\_add}}.
\Y\B\4\X12:Has \PB{\\{vf}} been described?\X${}\E{}$\6
\&{if} ${}(\\{vf}\MG\|N<\T{0}){}$\5
${}\{{}$\1\6
${}\\{LogError}(\.{EC\_ERROR},\39\.{VFN},\39\.{"Voting\ farm\ \%d\ need}\)\.{s\
to\ be\ described."},\39\\{vf}\MG\\{vf\_id});{}$\6
${}\\{LogError}(\.{EC\_ERROR},\39\.{VFN},\39\.{"\\t(You\ probably\ nee}\)\.{d\
to\ execute\ a\ VF\_ad}\)\.{d\ statement.)"});{}$\6
\&{return} \\{VF\_error}${}\K\.{E\_VF\_UNDESCRIBED};{}$\6
\4${}\}{}$\2\par
\Us11\ET24.\fi

\M{13}This checks whether the farm has been activated by means of
a previous call to function \PB{\\{VF\_run}}.
\Y\B\4\X13:Has \PB{\\{vf}} been activated?\X${}\E{}$\6
\&{if} ${}(\\{vf}\MG\\{user\_thread}<\T{0}){}$\5
${}\{{}$\1\6
${}\\{LogError}(\.{EC\_ERROR},\39\.{VFN},\39\.{"Voting\ farm\ \%d\ need}\)\.{s\
to\ be\ activated."},\39\\{vf}\MG\\{vf\_id});{}$\6
${}\\{LogError}(\.{EC\_ERROR},\39\.{VFN},\39\.{"\\t(You\ probably\ nee}\)\.{d\
to\ execute\ a\ VF\_ru}\)\.{n\ statement.)"});{}$\6
\&{return} \\{VF\_error}${}\K\.{E\_VF\_INACTIVE};{}$\6
\4${}\}{}$\2\par
\U24.\fi

\N{1}{14}Voting Farm Control.
All interactions between the user module and the farm go through
the \PB{\\{VF\_control}} and \PB{\\{VF\_control\_list}} functions and objects
of \PB{$\VFxmsgxt$} type,
aka messages.

\Y\B\4\X14:Voting Farm Control\X${}\E{}$\6
\X15:Type \PB{$\VFxmsgxt$}\X\6
\X17:Build a \PB{$\VFxmsgxt$} message\X\6
\X24:Function \PB{\\{VF\_control\_list}}\X\6
\X26:Function \PB{\\{VF\_control}}\X\6
\X25:Function \PB{\\{VF\_send}}\X\par
\U1.\fi

\M{15}A message is an object which holds the information needed
for a user module to request a service to a voter and for a voter to
respond to a previous user's request.
Its definition is simple:

\Y\B\4\X15:Type \PB{\&{VF\_msg\_t}}\X${}\E{}$\6
\&{typedef} \&{struct} ${}\{{}$\1\6
\&{int} \\{code};\6
\&{void} ${}{*}\\{msg};{}$\6
\&{int} \\{msglen};\2\6
${}\}{}$ ${}\&{VF\_msg\_t}{}$;\par
\U14.\fi

\M{16}The \PB{\\{code}} field specifies the nature of the message (see table~%
\ref{table1}
for a complete reference); depending on this, some data may be pointed to
by the opaque pointer \PB{\\{msg}}. In this case, \PB{\\{msglen}} represents
the size of that data. This is an example of its usage:

\vskip 5pt

\item{}\hskip 35pt{\bf (1)} \PB{\&{VF\_msg\_t} \\{message};}
\item{}\hskip 35pt{\bf (2)} \PB{$\\{message}.\\{code}\K\.{VF\_INP\_MSG};$}
\item{}\hskip 35pt{\bf (3)} \PB{$\\{message}.\\{msg}\K\\{strdup}(\\{input});$}
\item{}\hskip 35pt{\bf (4)} \PB{$\\{message}.\\{msglen}\K\T{1}+\\{strlen}(%
\\{input});$}

\vskip 5pt

\noindent
Note that the voter thread assumes that the user module allocates
new memory for each new \PB{$\\{message}.\\{msg}$} that is sent to it i.e.,
$\underline{\hbox{\it it won't make a personal copy of the area\/}}$;
on the contrary, it will simply
store the pointer and use the pointed storage.
Moreover, also deallocation of objects previously defined
by the user module will be considered to be managed by this latter.

\fi

\M{17}These functions hide the \PB{\&{VF\_msg\_t}} structure to the user.
\Y\B\4\X17:Build a \PB{\&{VF\_msg\_t}} message\X${}\E{}$\6
\X18:Input message setup\X\6
\X19:Scaling factor message setup\X\6
\X20:Message to choose the algorithm\X\par
\U14.\fi

\M{18}Build up an \PB{\&{VF\_msg\_t}} object holding a \PB{\.{VF\_INP\_MSG}}
message.
\Y\B\4\X18:Input message setup\X${}\E{}$\6
\&{VF\_msg\_t} ${}{*}{}$\\{VFO\_Set\_Input\_Message}(\&{void} ${}{*}\\{obj},%
\39{}$\&{size\_t} \\{siz})\1\1\2\2\6
${}\{{}$\1\6
\&{static} \&{VF\_msg\_t} \|m;\7
\&{if} ${}(\\{obj}\E\NULL){}$\5
${}\{{}$\1\6
${}\\{VF\_error}\K\.{E\_VF\_NULLPTR};{}$\6
\&{return} ${}\NULL;{}$\6
\4${}\}{}$\2\6
${}\|m.\\{code}\K\.{VF\_INP\_MSG};{}$\6
${}\|m.\\{msg}\K\\{obj};{}$\6
${}\|m.\\{msglen}\K\\{siz};{}$\6
\&{return} ${}{\AND}\|m;{}$\6
\4${}\}{}$\2\par
\U17.\fi

\M{19}Build up an \PB{\&{VF\_msg\_t}} object holding a \PB{\.{VF\_SCALING}}
message.
\Y\B\4\X19:Scaling factor message setup\X${}\E{}$\6
\&{VF\_msg\_t} ${}{*}{}$\\{VFO\_Set\_Scaling\_Factor}(\&{double} ${}{*}%
\\{sf}){}$\1\1\2\2\6
${}\{{}$\1\6
\&{static} \&{VF\_msg\_t} \|m;\7
${}\|m.\\{code}\K\.{VF\_SCALING\_FACTOR};{}$\6
${}\|m.\\{msg}\K\\{sf};{}$\6
${}\|m.\\{msglen}\K\&{sizeof}(\&{double});{}$\6
\&{return} ${}{\AND}\|m;{}$\6
\4${}\}{}$\2\par
\U17.\fi

\M{20}Build up an \PB{\&{VF\_msg\_t}} object holding the chosen algorithm.
\Y\B\4\X20:Message to choose the algorithm\X${}\E{}$\6
\&{VF\_msg\_t} ${}{*}{}$\\{VFO\_Set\_Algorithm}(\&{int} \\{algorithm})\1\1\2\2\6
${}\{{}$\1\6
\&{static} \&{VF\_msg\_t} \|m;\7
${}\|m.\\{code}\K\.{VF\_SELECT\_ALG};{}$\6
${}\|m.\\{msglen}\K\\{algorithm};{}$\6
\&{return} ${}{\AND}\|m;{}$\6
\4${}\}{}$\2\par
\U17.\fi

\M{21}Function \PB{\\{VF\_control\_list}} accepts an array of messages that are
transferred
across the communication network as one buffer.

Function \PB{\\{VF\_control\_list}} and \PB{\\{VF\_control}} return a negative
integer in case of error; otherwise,
they return zero.

\fi

\M{22}List of possible message codes going: from the user module to its
local voter (represented as
${\cal U}\rightarrow{\cal L}$),
and vice-versa (represented as
${\cal U}\rightarrow{\cal L}$):

\item{$\bullet$} \PB{\.{VF\_INP\_MSG}}:
a \PB{\\{msglen}}-byte-long input message
is stored at the address referenced by the opaque pointer \PB{\\{msg}}
(${\cal U}\rightarrow{\cal L}$).
\item{$\bullet$} \PB{\.{VF\_OUT\_LCB}}:
the information pointed to by \PB{\\{msg}} is the link control
block for connecting the local voter to the output module
(${\cal U}\rightarrow{\cal L}$); see Figure 3.
\item{$\bullet$} \PB{\.{VF\_SELECT\_ALG}}:
\PB{\\{msg}} points to a code which identifies a particular
majority voting algorithm out of the set of available algorithms
(${\cal U}\rightarrow{\cal L}$).
\item{$\bullet$} \PB{\.{VF\_DESTROY}}:
a signal meaning that the receiving voter should
terminate itself after having freed all no more needed memory
(${\cal U}\rightarrow{\cal L}$).
\item{$\bullet$} \PB{\.{VF\_NOP}}:
\.{N}o \.{OP}eration, something like an \.{ImAlive} signal.
For the time being this event is not used.
\item{$\bullet$} \PB{\.{VF\_RESET}}:
on the arrival of this segnal the status is set so to be able
to perform a new voting session with the current farm
(${\cal U}\rightarrow{\cal L}$).
\item{$\bullet$} \PB{\.{VF\_REFUSED}}:
certain operations may be refused; for example, if one
tries to execute a \PB{\\{VF\_close}(\,)} before a broadcasting
operation has been completed, then the voter returns
this message to its user module
(${\cal L}\rightarrow{\cal U}$).
\item{$\bullet$} \PB{\.{VF\_QUIT}}:
before quitting, the voter generates a \PB{\.{VF\_QUIT}} event
(${\cal L}\rightarrow{\cal U}$).
\item{$\bullet$} \PB{\.{VF\_DONE}}:
after broadcasting, a \PB{\.{VF\_DONE}} event is raised
(${\cal L}\rightarrow{\cal U}$).
\item{$\bullet$} \PB{\.{VF\_EPSILON}}:
An $\epsilon$ threshold value needed by the formalized
majority voting algorithm
(${\cal U}\rightarrow{\cal L}$).
\item{$\bullet$} \PB{\.{VF\_ERROR}}:
A generic error has occurred
(${\cal L}\rightarrow{\cal U}$).
\item{$\bullet$} \PB{\.{VF\_SCALING\_FACTOR}}:
used in the Weighted Averaging Technique (see below)
(${\cal U}\rightarrow{\cal L}$).

\vskip 5pt

\centerline{\psfig{file=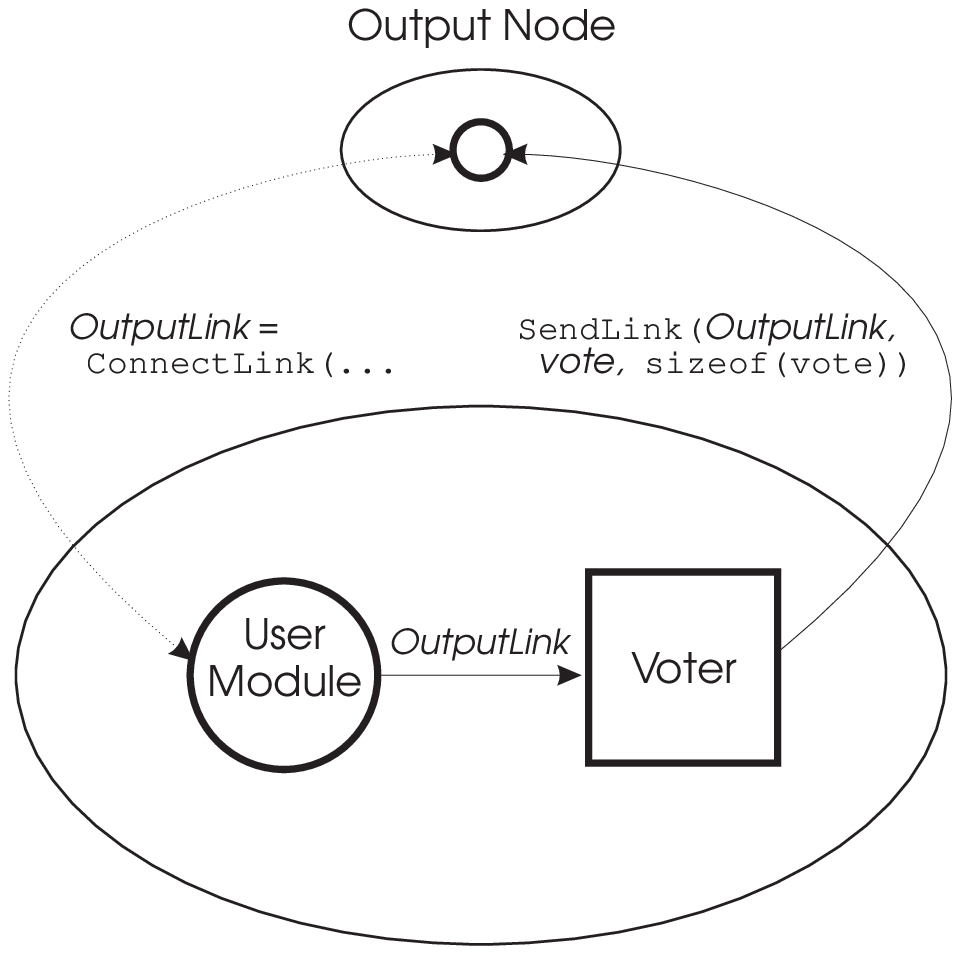}}
\centerline{{\bf Figure 3.}
\ninerm
The user module connects to an output module (dashed curve) and then sends to
its}
\vskip -3pt
\centerline{\ninerm
local voter a \PB{\.{VF\_OUT\_LCB}} message holding the relevant link control.
As soon as
the voter has perf-}
\vskip -3pt
\centerline{\ninerm
ormed its voting task, it sends its vote to the output module
by means of that link control block.}

\vskip 5pt

\Y\B\4\D$\.{VF\_INP\_MSG}$ \5
\T{100}\par
\B\4\D$\.{VF\_OUT\_LCB}$ \5
\T{101}\par
\B\4\D$\.{VF\_SELECT\_ALG}$ \5
\T{102}\par
\B\4\D$\.{VF\_DESTROY}$ \5
\T{103}\par
\B\4\D$\.{VF\_NOP}$ \5
\T{104}\par
\B\4\D$\.{VF\_RESET}$ \5
\T{105}\par
\B\4\D$\.{VF\_REFUSED}$ \5
\T{106}\par
\B\4\D$\.{VF\_QUIT}$ \5
\T{107}\par
\B\4\D$\.{VF\_DONE}$ \5
\T{108}\par
\B\4\D$\.{VF\_EPSILON}$ \5
\T{109}\par
\B\4\D$\.{VF\_ERROR}$ \5
\T{110}\par
\B\4\D$\.{VF\_SCALING\_FACTOR}$ \5
\T{111}\par
\fi

\M{23}List of possible message codes going from the local voter to its
fellows and vice-versa. They have the same meaning of those defined
in the previous section e.g., \PB{\.{VF\_V\_INP\_MSG}} is an input message
coming from a fellow voter.

\Y\B\4\D$\.{VF\_V\_INP\_MSG}$ \5
\T{200}\par
\B\4\D$\.{VF\_V\_DESTROY}$ \5
\T{203}\par
\B\4\D$\.{VF\_V\_NOP}$ \5
\T{204}\par
\B\4\D$\.{VF\_V\_RESET}$ \5
\T{205}\par
\B\4\D$\.{VF\_V\_ERROR}$ \5
\T{210}\par
\fi

\M{24}This function sends one or more messages to the local voter
of voting farm \PB{\\{vf}}:

\Y\B\4\X24:Function \PB{\\{VF\_control\_list}}\X${}\E{}$\6
\&{int} \\{VF\_control\_list}(\&{VotingFarm\_t} ${}{*}\\{vf},\39{}$\&{VF\_msg%
\_t} ${}{*}\\{msg},\39{}$\&{int} \|n)\1\1\2\2\6
${}\{{}$\1\6
\\{AllocationClass}\1\1\6
\&{int} \\{rv};\C{ set function name }\6
\&{static} \&{char} ${}{*}\.{VFN}\K\.{"VF\_control\_list"};\2\2{}$\6
\8\#\&{ifdef} \.{VFDEBUG}\7
${}\\{LogError}(\.{EC\_MESS},\39\.{VFN},\39\.{"within\ control\_list}\)%
\.{"});{}$\6
\8\#\&{endif}\C{ VFDEBUG }\7
\X8:Has \PB{\\{vf}} been defined?\X\6
\X12:Has \PB{\\{vf}} been described?\X\6
\X13:Has \PB{\\{vf}} been activated?\X\7
\8\#\&{ifdef} \.{VFDEBUG}\6
${}\\{LogError}(\.{EC\_MESS},\39\.{VFN},\39\.{"checked:\ defined,\ d}\)%
\.{escribed,\ and\ activa}\)\.{ted"});{}$\6
\8\#\&{endif}\C{ VFDEBUG }\7
\&{if} ${}(\|n<\T{1}\V\|n>\.{VF\_MAX\_MSGS}){}$\5
${}\{{}$\C{ doubt: should it be an error or a warning? }\1\6
${}\\{LogError}(\.{EC\_ERROR},\39\.{VFN},\39\.{"Wrong\ message\ numbe}\)\.{r\ (%
\%d)\ ---\ should\ be}\)\.{\ between\ 1\ and\ \%d.\\n}\)\.{"},\39\|n,\39\.{VF%
\_MAX\_MSGS});{}$\6
\&{return} \.{E\_VF\_WRONG\_MSG\_NB};\6
\4${}\}{}$\2\6
\&{if} ${}(\\{vf}\MG\\{pipe}[\T{0}]\I\NULL){}$\5
${}\{{}$\6
\8\#\&{ifdef} \.{VFDEBUG}\1\6
${}\\{LogError}(\.{EC\_MESS},\39\.{VFN},\39\.{"Next\ statement\ is\ S}\)%
\.{endLink(\%x,\ \%x,\ \%d)"},\39\\{vf}\MG\\{pipe}[\T{0}],\39{\AND}\\{msg}[%
\T{0}],\39\|n*{}$\&{sizeof} (\\{msg}[\T{0}]));\6
\8\#\&{endif}\C{ VFDEBUG }\7
${}\\{rv}\K\\{SendLink}(\\{vf}\MG\\{pipe}[\T{0}],\39{\AND}\\{msg}[\T{0}],\39%
\|n*{}$\&{sizeof} (\\{msg}[\T{0}]));\6
\&{if} ${}(\\{rv}\I\|n*{}$\&{sizeof} (\\{msg}[\T{0}]))\5
${}\{{}$\1\6
${}\\{LogError}(\.{EC\_ERROR},\39\.{VFN},\39\.{"Cannot\ send\ the\ mes}\)%
\.{sage\ to\ local\ voter.}\)\.{"});{}$\6
\&{return} \.{E\_VF\_SENDLINK};\6
\4${}\}{}$\2\6
\8\#\&{ifdef} \.{VFDEBUG}\6
${}\\{LogError}(\.{EC\_MESS},\39\.{VFN},\39\.{"exiting..."});{}$\6
\8\#\&{endif}\C{ VFDEBUG }\6
\&{return} \\{VF\_error}${}\K\T{0};{}$\6
\4${}\}{}$\2\6
${}\\{LogError}(\.{EC\_ERROR},\39\.{VFN},\39\.{"Corrupted\ or\ Invali}\)\.{d\
VotingFarm\_t\ Objec}\)\.{t."});{}$\6
\&{return} \\{VF\_error}${}\K\.{E\_VF\_INVALID\_VF};{}$\6
\4${}\}{}$\2\par
\U14.\fi

\M{25}To be described...
\Y\B\4\X25:Function \PB{\\{VF\_send}}\X${}\E{}$\6
\8\#\&{define} \.{VF\_MAXARGS} \5\T{31}\6
\&{void} \\{VF\_send}(\&{VotingFarm\_t} ${}{*}\\{vf},\39{}$\&{int} \\{argc}${},%
\39\,\ldots\,){}$\1\1\2\2\6
${}\{{}$\1\6
\\{AllocationClass}\1\1\6
\&{va\_list} \\{ap};\2\2\6
\\{AllocationClass}\1\1\6
\&{VF\_msg\_t} \\{array}[\.{VF\_MAXARGS}];\2\2\6
\\{AllocationClass}\1\1\6
\&{VF\_msg\_t} ${}{*}\\{mp};\2\2{}$\6
\\{AllocationClass}\1\1\6
\&{int} \|i;\2\2\6
\&{if} ${}(\\{argc}>\.{VF\_MAXARGS}){}$\1\5
${}\\{argc}\K\.{VF\_MAXARGS};{}$\2\7
${}\\{va\_start}(\\{ap},\39\\{argc});{}$\7
\&{for} ${}(\|i\K\T{0};{}$ ${}\|i<\\{argc};{}$ ${}\|i\PP){}$\5
${}\{{}$\1\6
${}\\{mp}\K\\{va\_arg}(\\{ap},\39{}$\&{VF\_msg\_t} ${}{*});{}$\6
${}\\{memcpy}(\\{array}+\|i,\39\\{mp},\39\&{sizeof}(\&{VF\_msg\_t}));{}$\6
\4${}\}{}$\2\7
\\{va\_end}(\\{ap});\7
${}\\{VF\_control\_list}(\\{vf},\39\\{array},\39\\{argc});{}$\6
\4${}\}{}$\2\par
\U14.\fi

\M{26}Simply a shortcut.
\Y\B\4\X26:Function \PB{\\{VF\_control}}\X${}\E{}$\6
\&{int} \\{VF\_control}(\&{VotingFarm\_t} ${}{*}\\{vf},\39{}$\&{VF\_msg\_t}
${}{*}\\{msg}){}$\1\1\2\2\6
${}\{{}$\1\6
\&{return} \\{VF\_control\_list}${}(\\{vf},\39\\{msg},\39\T{1});{}$\6
\4${}\}{}$\2\par
\U14.\fi

\M{27}Voting Farm Destruction.
This is another simple shortcut for sending the \PB{\.{VF\_DESTROY}} message
to a voting farm.

\Y\B\4\X27:Voting Farm Destruction\X${}\E{}$\6
\&{int} \\{VF\_close}(\&{VotingFarm\_t} ${}{*}\\{vf}){}$\1\1\2\2\6
${}\{{}$\1\6
\\{AllocationClass}\1\1\6
\&{VF\_msg\_t} \\{msg};\2\2\6
\\{AllocationClass}\1\1\6
\&{int} \|r;\2\2\6
${}\\{msg}.\\{code}\K\.{VF\_DESTROY};{}$\7
${}\|r\K\\{VF\_control}(\\{vf},\39{\AND}\\{msg});{}$\7
\&{return} \|r;\6
\4${}\}{}$\2\par
\U1.\fi

\M{28}to be described.
\Y\B\4\X28:Voting Farm Read\X${}\E{}$\6
\&{VF\_msg\_t} ${}{*}{}$\\{VF\_get}(\&{VotingFarm\_t} ${}{*}\\{vf}){}$\1\1\2\2\6
${}\{{}$\1\6
\&{static} \&{VF\_msg\_t} \\{msg};\7
\\{AllocationClass}\1\1\6
\&{int} \\{recv}${},{}$ \|n;\2\2\6
${}\\{AllocationClass}\Optionxt\|o[\T{5}]{}$;\C{ used to be \PB{$\Optionxt\|o[%
\T{2}]$} }\7
\&{static} \&{char} ${}{*}\.{VFN}\K\.{"VF\_get"};{}$\6
\8\#\&{ifdef} \.{VFDEBUG}\7
${}\\{LogError}(\.{EC\_MESS},\39\.{VFN},\39\.{"\ vf==\%x,\ vf->pipe==}\)\.{\%x,%
\ vf->pipe[0]==\%x,}\)\.{\ vf->pipe[1]==\%x"},\39\\{vf},\39\\{vf}\MG\\{pipe},%
\39\\{vf}\MG\\{pipe}[\T{0}],\39\\{vf}\MG\\{pipe}[\T{1}]);{}$\6
\8\#\&{endif}\C{ VFDEBUG }\6
${}\|o[\T{0}]\K\\{ReceiveOption}(\\{vf}\MG\\{pipe}[\T{0}]);{}$\6
${}\|o[\T{1}]\K\\{TimeAfterOption}(\\{TimeNow}(\,)+\.{VF\_EVENT\_TIMEOUT}*%
\.{CLOCK\_TICK}){}$;\7
\&{switch} ${}((\|n\K\\{SelectList}(\T{2},\39\|o))){}$\5
${}\{{}$\1\6
\4\&{case} \T{1}:\5
${}\\{LogError}(\.{EC\_ERROR},\39\.{VFN},\39\.{"Timeout\ condition\ r}\)%
\.{eached."});{}$\6
${}\\{LogError}(\.{EC\_ERROR},\39\.{VFN},\39\.{"\\t(Maybe\ VF\_EVENT\_T}\)%
\.{IMEOUT\ should\ be\ enl}\)\.{arged?\ Now\ it's\ \%d\ s}\)\.{econds.)"},\39%
\.{VF\_EVENT\_TIMEOUT});{}$\6
${}\\{msg}.\\{code}\K\.{VF\_ERROR};{}$\6
${}\\{msg}.\\{msglen}\K\\{VF\_error}\K\.{E\_VF\_EVENT\_TIMEOUT};{}$\6
\&{return} ${}{\AND}\\{msg};{}$\6
\4\&{case} \T{0}:\6
\&{if} ${}((\\{recv}\K\\{RecvLink}(\\{vf}\MG\\{pipe}[\T{0}],\39{}$(\&{void}
${}{*}){}$ ${}{\AND}\\{msg},\39{}$\&{sizeof} (\\{msg})))${}\Z\T{0}){}$\5
${}\{{}$\1\6
${}\\{LogError}(\.{EC\_ERROR},\39\.{VFN},\39\.{"Can't\ RecvLink\ [A\ m}\)%
\.{essage\ from\ the\ loca}\)\.{l\ voter]"});{}$\6
${}\\{LogError}(\.{EC\_ERROR},\39\.{VFN},\39\.{"\\t(return\ code\ is\ \%}\)%
\.{d.)"},\39\\{recv});{}$\6
${}\\{msg}.\\{code}\K\.{VF\_ERROR};{}$\6
${}\\{msg}.\\{msglen}\K\\{VF\_error}\K\.{E\_VF\_RECVLINK};{}$\6
\&{return} ${}{\AND}\\{msg};{}$\6
\4${}\}{}$\2\6
\&{break};\6
\4\&{default}:\5
${}\\{LogError}(\.{EC\_ERROR},\39\.{VFN},\39\.{"Can't\ Select\ ---\ re}\)%
\.{tvalue\ is\ \%d"},\39\|n);{}$\6
${}\\{msg}.\\{code}\K\.{VF\_ERROR};{}$\6
${}\\{msg}.\\{msglen}\K\\{VF\_error}\K\.{E\_VF\_SELECT};{}$\6
\&{return} ${}{\AND}\\{msg};{}$\6
\4${}\}{}$\C{ end \PB{\&{switch}} }\2\7
\8\#\&{ifdef} \.{VFDEBUG}\6
${}\\{LogError}(\.{EC\_MESS},\39\.{VFN},\39\.{"Received\ a\ message\ }\)\.{(\%d%
\ bytes)\ from\ the\ }\)\.{voter."},\39\\{recv});{}$\6
\8\#\&{endif}\C{ VFDEBUG }\6
\&{return} ${}{\AND}\\{msg};{}$\6
\4${}\}{}$\2\7
\&{static} \&{int} ${}\\{voter\_sendcode}(\LinkCBxt*\\{UserLink},\39{}$\&{int} %
\\{code})\1\1\2\2\6
${}\{{}$\1\6
\\{AllocationClass}\1\1\6
\&{int} \\{rv};\2\2\6
\\{AllocationClass}\1\1\6
\&{VF\_msg\_t} \\{msg};\6
\&{static} \&{char} ${}{*}\.{VFN}\K\.{"voter\_sendcode"};\2\2{}$\6
${}\\{msg}.\\{code}\K\\{code};{}$\7
${}\\{rv}\K\\{SendLink}(\\{UserLink},\39{\AND}\\{msg},\39{}$\&{sizeof} (%
\\{msg}));\7
\&{if} ${}(\\{rv}\I{}$\&{sizeof} (\\{msg}))\5
${}\{{}$\1\6
${}\\{LogError}(\.{EC\_ERROR},\39\.{VFN},\39\.{"Cannot\ send\ the\ mes}\)%
\.{sage\ to\ the\ user\ mod}\)\.{ule."});{}$\6
\&{return} \.{E\_VF\_SENDLINK};\6
\4${}\}{}$\2\6
\&{return} \\{rv};\6
\4${}\}{}$\2\7
\&{static} \&{int} ${}\\{voter\_sendmsg}(\LinkCBxt*\\{UserLink},\39{}$\&{VF%
\_msg\_t} ${}{*}\\{msg}){}$\1\1\2\2\6
${}\{{}$\1\6
\\{AllocationClass}\1\1\6
\&{int} \\{rv};\6
\&{static} \&{char} ${}{*}\.{VFN}\K\.{"voter\_sendmsg"}{}$;\C{ \PB{\&{static}}
function $\Rightarrow$ \PB{\.{VFN}} is not touched }\2\2\6
${}\\{rv}\K\\{SendLink}(\\{UserLink},\39\\{msg},\39{}$\&{sizeof} ${}({*}%
\\{msg}));{}$\7
\&{if} ${}(\\{rv}\I{}$\&{sizeof} ${}({*}\\{msg})){}$\5
${}\{{}$\1\6
${}\\{LogError}(\.{EC\_ERROR},\39\.{VFN},\39\.{"Cannot\ send\ the\ mes}\)%
\.{sage\ to\ the\ user\ mod}\)\.{ule."});{}$\6
\&{return} \.{E\_VF\_SENDLINK};\6
\4${}\}{}$\2\6
\&{return} \\{rv};\6
\4${}\}{}$\2\par
\U1.\fi

\M{29}Voting functions are stored into the \PB{\\{DoVoting}} array.
\Y\B\4\X29:Voting Algorithms\X${}\E{}$\6
\&{typedef} \&{struct} ${}\{{}$\1\6
\&{unsigned} \&{char} ${}{*}\\{vote};{}$\6
\&{int} \\{outcome};\2\6
${}\}{}$ \&{vote\_t};\7
\X52:Voting Functions\X\7
\&{typedef} \&{void} ${}({*}\&{voter\_t}){}$(\&{VotingFarm\_t} ${}{*},\39{}$%
\&{void} ${}{*}[\,],\39\&{int},\39{}$\&{vote\_t} ${}{*}){}$;\6
\&{static} \&{voter\_t} \\{DoVoting}[\.{VF\_NB\_ALGS}]${}\K\{\\{VFA%
\_ExactConcensus},{}$\6
\\{VFA\_MajorityVoting}${},{}$\6
\\{VFA\_MedianVoting}${},{}$\6
\\{VFA\_PluralityVoting}${},{}$\6
\\{VFA\_WeightedAveraging}${},{}$\6
\\{VFA\_SimpleMajorityVoting}${},{}$\6
\\{VFA\_SimpleAverage}${},{}$\6
${}\}{}$;\par
\U1.\fi

\N{1}{30}The Voter Function. The function at the basis of the voting thread.

\Y\B\4\X30:The Voter Function\X${}\E{}$\6
\&{static} \&{int} \\{VF\_voter}(\&{VotingFarm\_t} ${}{*}\\{vf}){}$\1\1\2\2\6
${}\{{}$\1\6
${}\LinkCBxt*\\{UserLink},\39{*}\\{OutputLink},\39{*}{*}\\{links};{}$\6
\8\#\&{ifdef} \.{SERVERNET}\6
${}\LinkCBxt*\\{serverlink}\K\\{ConnectServer}(\,);{}$\6
\8\#\&{endif}\C{ SERVERNET }\6
${}\Optionxt*\\{options};{}$\7
\&{int} \\{this\_voter}${},{}$ \\{this\_node}${},{}$ \|N${},{}$ \|i${},{}$ %
\\{opt}${},{}$ \\{input\_nr};\6
\&{VF\_msg\_t} \\{msg}[\.{VF\_MAX\_MSGS}];\6
\&{int} \\{msgnum};\6
\&{VF\_msg\_t} \\{done};\6
\&{int} \\{Algorithm}${}\K\.{VFA\_MAJORITY};{}$\6
\&{char} ${}\\{buffer}[\.{VF\_MAX\_INPUT\_MSG}+\&{sizeof}(\&{int})];{}$\6
\&{int} \\{Error}${},{}$ \\{recv};\6
\&{int} \\{input\_length};\6
\&{void} ${}{*}{*}\\{voter\_inputs};{}$\6
\&{unsigned} \&{char} ${}{*}\\{vote}\K\\{st\_VFA\_vote};{}$\6
\&{vote\_t} \\{rvote};\6
\&{unsigned} \&{int} \\{t0}${},{}$ \\{tn};\C{ set function name }\6
\&{static} \&{char} ${}{*}\.{VFN}\K\.{"VF\_voter"};{}$\7
${}\\{t0}\K\\{TimeNow}(\,);{}$\6
\8\#\&{ifdef} \.{SERVERNET}\6
${}\\{Set\_Phase}(\\{serverlink},\39\.{VFP\_INITIALISING});{}$\6
\8\#\&{endif}\6
${}\\{input\_nr}\K\\{input\_length}\K\T{0};{}$\6
${}\\{UserLink}\K\\{OutputLink}\K\NULL;{}$\6
${}\\{vf}\MG\\{broadcast\_done}\K\\{vf}\MG\\{inp\_msg\_got}\K\\{vf}\MG%
\\{destroy\_requested}\K\.{NO};{}$\6
${}\\{rvote}.\\{vote}\K\\{vote}{}$;\7
\X8:Has \PB{\\{vf}} been defined?\X\6
\X9:Is \PB{\\{vf}} a valid object?\X\7
${}\\{this\_voter}\K\\{vf}\MG\\{this\_voter};{}$\6
${}\\{this\_node}\K\\{vf}\MG\\{vf\_node\_stack}[\\{this\_voter}];{}$\6
\&{if} ${}(\.{GET\_ROOT}(\,)\MG\\{ProcRoot}\MG\\{MyProcID}\I\\{this\_node}){}$\5
${}\{{}$\1\6
${}\\{LogError}(\.{EC\_ERROR},\39\.{VFN},\39\.{"Corrupted\ or\ Invali}\)\.{d\
VotingFarm\_t\ Objec}\)\.{t."});{}$\6
\&{return} \\{VF\_error}${}\K\.{E\_VF\_INVALID\_VF};{}$\6
\4${}\}{}$\2\6
${}\|N\K\\{vf}\MG\|N{}$;\C{ Connect to the local Agent }\6
\8\#\&{ifdef} \.{SERVERNET}\6
\X61:Ask the Server to set up a connection to an Agent\X\6
\8\#\&{endif}\C{ SERVERNET }\C{ Get the \PB{( $\LinkCBxt$ $*$ )} for
communicating with the user module }\6
${}\\{UserLink}\K\\{vf}\MG\\{pipe}[\T{1}]{}$;\7
\X32:Create a cliqu\'e\X\6
\X37:Poll the user link, the farm links, and the server link\X\6
\4${}\}{}$\2\6
\X31:An example of metric function\X\par
\U1.\fi

\M{31}An example of metric function: a double-returning
version of \PB{\\{strcmp}}.
\Y\B\4\X31:An example of metric function\X${}\E{}$\6
\&{double} \\{dstrcmp}(\&{void} ${}{*}\|a,\39{}$\&{void} ${}{*}\|b){}$\1\1\2\2\6
${}\{{}$\1\6
\&{return} (\&{double}) \\{strcmp}${}(\|a,\39\|b);{}$\6
\4${}\}{}$\2\par
\U30.\fi

\M{32}A cliqu\'e (fully interconnected crossbar) is set up among the voters.
This is done in two ``flavours:'' in the \PB{\.{STATIC}} one we make use of
predefined static data, otherwise we perform \PB{\\{malloc}(\,)}'s.
\Y\B\4\X32:Create a cliqu\'e\X${}\E{}$\6
\8\#\&{ifndef} \.{STATIC}\6
\X33:Allocate an array of \PB{$\LinkCBxt$} pointers\X\6
\8\#\&{else}\6
\X34:Initialize an array of \PB{$\LinkCBxt$} pointers\X\6
\8\#\&{endif}\C{ \PB{\.{STATIC}} }\6
\X35:Connect to your \PB{\|N}-1 fellows\X\par
\U30.\fi

\M{33}An array of Link Control Block Pointers is needed in order to
realize the cliqu\'e.
\Y\B\4\X33:Allocate an array of \PB{$\LinkCBxt$} pointers\X${}\E{}$\6
\\{links} $\K$ ( $\LinkCBxt$ $*$ $*$ ) \\{malloc} ( \|N $*$ \&{sizeof} ( $%
\LinkCBxt$ $*$ ) )  ; \\{options} $\K$ ( $\Optionxt$ $*$ ) $\\{malloc}((\|N+%
\T{1})*{}$\&{sizeof} ${}(\Optionxt));{}$\6
${}\\{voter\_inputs}\K{}$(\&{void} ${}{*}{*}){}$ \\{calloc}${}(\|N,\39{}$%
\&{sizeof}(\&{void} ${}{*}));{}$\6
\&{if} ${}(\\{links}\E\NULL\V\\{options}\E\NULL\V\\{voter\_inputs}\E\NULL){}$\5
${}\{{}$\1\6
${}\\{LogError}(\.{EC\_ERROR},\39\.{VFN},\39\.{"Memory\ Allocation\ E}\)%
\.{rror."});{}$\6
\&{return} \\{VF\_error}${}\K\.{E\_VF\_CANT\_ALLOC};{}$\6
\4${}\}{}$\2\par
\U32.\fi

\M{34}This section is supplied in case the user is willing to use the {\it
static\/}
version of this tool. Rather than allocating memory, we link
statically-allocated memory to the appropriate pointers and we initialize
them---where needed.
\Y\B\4\X34:Initialize an array of \PB{$\LinkCBxt$} pointers\X${}\E{}$\6
$\\{links}\K\\{st\_links};{}$\6
${}\\{options}\K\\{st\_options};{}$\6
${}\\{voter\_inputs}\K\\{st\_voter\_inputs};{}$\6
${}\\{memset}(\\{voter\_inputs},\39\T{0},\39\.{VF\_STATIC\_MAX\_INP\_MSG}){}$;%
\par
\U32.\fi

\M{35}``By the implementation of queues the calling order $\{$to \PB{%
\\{ConnectLink}}$\}$
does not matter.'' [{\sc 11}]. As a consequence,
a simple \PB{\&{for}} should suffice to create the farm.

\PB{\|N}+1 options are ``received'': \PB{\|N}-1 for the cliqu\'e set-up, 1 for
connecting to the user module, 1 for connecting to the server; more
precisely:

\item{$\bullet$}{\PB{\\{options}[\\{this\_voter}]}} regards the user module;
\item{$\bullet$}{\PB{\\{options}[\|N]}} regards the local server;
\item{$\bullet$}{the rest} regards the fellow voters.

\Y\B\4\X35:Connect to your \PB{\|N}-1 fellows\X${}\E{}$\6
\8\#\&{ifdef} \.{SERVERNET}\6
$\\{Set\_Phase}(\\{serverlink},\39\.{VFP\_CONNECTING});{}$\6
\8\#\&{endif}\6
\&{for} ${}(\|i\K\T{0};{}$ ${}\|i<\|N;{}$ ${}\|i\PP){}$\5
${}\{{}$\1\6
\&{if} ${}(\|i\I\\{this\_voter}){}$\5
${}\{{}$\1\6
${}\\{links}[\|i]\K\\{ConnectLink}(\\{vf}\MG\\{vf\_node\_stack}[\|i],\39\\{VF%
\_RequestId}(\\{vf}\MG\\{vf\_id},\39\|i,\39\\{this\_voter}),\39{\AND}%
\\{Error});{}$\6
${}\\{options}[\|i]\K\\{ReceiveOption}(\\{links}[\|i]);{}$\6
\&{if} ${}(\\{links}[\|i]\E\NULL){}$\5
${}\{{}$\1\6
${}\\{LogError}(\.{EC\_ERROR},\39\.{VFN},\39\.{"Cannot\ connect\ to\ v}\)%
\.{oter\ \%d."},\39\|i);{}$\6
\&{return} \\{VF\_error}${}\K\.{E\_VF\_CANT\_CONNECT};{}$\6
\4${}\}{}$\2\6
\4${}\}{}$\2\6
\&{else}\5
${}\{{}$\1\6
${}\\{options}[\|i]\K\\{ReceiveOption}(\\{UserLink});{}$\6
\4${}\}{}$\2\6
\4${}\}{}$\2\6
\8\#\&{ifdef} \.{SERVERNET}\6
${}\\{options}[\|N]\K\\{ReceiveOption}(\\{serverlink});{}$\6
\8\#\&{endif}\C{ SERVERNET }\par
\U32.\fi

\M{36}This function creates a new request id beginning from
voting farm id \PB{\\{vfn}} and voter id's \PB{\|v} and \PB{\|w}.

\Y\B\&{static} \&{int} \\{VF\_RequestId}(\&{int} \\{vfn}${},\39{}$\&{int} %
\|v${},\39{}$\&{int} \|w)\1\1\2\2\6
${}\{{}$\1\6
\\{AllocationClass}\1\1\6
\&{int} \|a${},{}$ \|b;\6
\&{static} \&{int} \\{hundreds}${}\K\.{VF\_MAX\_NTS}*\.{VF\_MAX\_NTS};\2\2{}$\6
\&{if} ${}(\|v>\|w){}$\1\5
${}\|a\K\|w,\39\|b\K\|v;{}$\2\6
\&{else}\1\5
${}\|a\K\|v,\39\|b\K\|w;{}$\2\7
\&{return} \\{hundreds}${}*(\\{vfn}+\T{1})+\|a*\.{VF\_MAX\_NTS}+\|b;{}$\6
\4${}\}{}$\2\par
\fi

\M{37}If control reaches this section it means that the cliqu\'e has been
successfully established between the voters. All future actions depend
on the control and input messages of the user; therefore, the voter
puts itself into an ``endless'' loop waiting for new messages to arrive
and to be managed. This event-driven loop indeed resembles the one that
may be found in any X11 client to manage interactions with the keyboard,
the pointer, the display server, and so on.

\Y\B\4\X37:Poll the user link, the farm links, and the server link\X${}\E{}$\C{
t0 = TimeNow(); }\6
\&{while} (\T{1})\7
${}\{{}$\6
\8\#\&{ifdef} \.{SERVERNET}\1\6
${}\\{opt}\K\\{SelectList}(\|N+\T{1},\39\\{options});{}$\6
\8\#\&{else}\6
${}\\{opt}\K\\{SelectList}(\|N,\39\\{options});{}$\6
\8\#\&{endif}\C{ SERVERNET }\7
\X38:A message from the user module\X\6
\&{else}\1\5
\X43:A message from the cliqu\'e\X\2\6
\8\#\&{ifdef} \.{SERVERNET}\6
\&{else}\1\5
\X46:A message from the server module\X\2\6
\8\#\&{endif}\C{ SERVERNET }\6
\&{else}\5
${}\{{}$\1\6
${}\\{LogError}(\.{EC\_ERROR},\39\.{VFN},\39\.{"Unknown\ sender\\n"});{}$\6
\&{return} \\{VF\_error}${}\K\.{E\_VF\_UNKNOWN\_SENDER};{}$\6
\4${}\}{}$\2\6
\&{if} ${}(\\{input\_nr}\E\|N\W\\{vf}\MG\\{broadcast\_done}\E\.{YES}\W%
\\{once}){}$\5
${}\{{}$\1\6
\&{FILE} ${}{*}\|f;{}$\6
\&{char} \\{fname}[\T{80}];\7
${}\\{tn}\K\\{TimeNow}(\,);{}$\6
\8\#\&{ifdef} \.{TIMESTATS}\6
${}\\{sprintf}(\\{fname},\39\.{"overhead.\%d"},\39\\{this\_voter});{}$\6
${}\|f\K\\{fopen}(\\{fname},\39\.{"a+"});{}$\6
${}\\{fprintf}(\|f,\39\.{"\%u\ \%u\ \%lf\\n"},\39\\{t0},\39\\{tn},\39(%
\&{double})(\\{tn}-\\{t0})/\.{CLOCK\_TICK});{}$\6
\\{fclose}(\|f);\6
\8\#\&{endif}\6
${}\\{once}\K\T{0};{}$\6
\4${}\}{}$\2\6
\4${}\}{}$\2\par
\U30.\fi

\M{38}If the received options is option \# \PB{\\{this\_voter}}, then a message
is
coming from the user module.

\Y\B\4\X38:A message from the user module\X${}\E{}$\6
\&{if} ${}(\\{opt}\E\\{this\_voter}){}$\5
${}\{{}$\C{ Receives the data and checks whether it is an integral multiple
   of the message size or not. In that latter case, an error is issued.      }%
\1\6
\&{if} ${}((\\{msgnum}\K\\{RecvLink}(\\{UserLink},\39\\{msg},\39\.{VF\_MAX%
\_MSGS}*{}$\&{sizeof} ${}({*}\\{msg})))\MOD{}$\&{sizeof} ${}({*}\\{msg})){}$\5
${}\{{}$\1\6
${}\\{LogError}(\.{EC\_ERROR},\39\.{VFN},\39\.{"Can't\ RecvLink\ [A\ m}\)%
\.{essage\ from\ the\ user}\)\.{\ module]"});{}$\6
${}\\{LogError}(\.{EC\_ERROR},\39\.{VFN},\39\.{"size\ of\ the\ message}\)\.{\
is\ \%d,\ sizeof(*msg)}\)\.{\ is\ \%d,\ (1)\%(2)\ is\ \%}\)\.{d"},\39%
\\{msgnum},\39{}$\&{sizeof} ${}({*}\\{msg}),\39\\{msgnum}\MOD{}$\&{sizeof}
${}({*}\\{msg}));{}$\6
\&{return} \\{VF\_error}${}\K\.{E\_VF\_RECVLINK};{}$\6
\4${}\}{}$\2\6
${}\\{msgnum}\MRL{{/}{\K}}{}$\&{sizeof} ${}({*}\\{msg});{}$\6
\8\#\&{ifdef} \.{VFDEBUG}\6
${}\\{LogError}(\.{EC\_MESS},\39\.{VFN},\39\.{"<voter\ \%d>\ Received}\)\.{\ a\
msg\ from\ the\ user}\)\.{\ module\ "},\39\\{this\_node});{}$\6
${}\\{LogError}(\.{EC\_MESS},\39\.{VFN},\39\.{"(code==\%d).\ msgnum=}\)\.{=\%d.%
\ Starting\ User\ m}\)\.{sg\ management\\n"},\39\\{msg}[\T{0}].\\{code},\39%
\\{msgnum});{}$\6
\8\#\&{endif}\7
${}\{{}$\1\6
\&{int} \|i;\7
\X39:User message management\X\6
\4${}\}{}$\2\6
\4${}\}{}$\2\par
\U37.\fi

\M{39}A message from the user has been received. Deal with that message.
(Note that messages are buffered into the \PB{\\{msg}[\,]} array.)

\Y\B\4\X39:User message management\X${}\E{}$\6
\&{for} ${}(\|i\K\T{0};{}$ ${}\|i<\\{msgnum};{}$ ${}\|i\PP)$ $\{$ \&{void}
${}{*}{}$\\{memdup}(\&{void} ${}{*},\39\&{size\_t});{}$\6
\8\#\&{ifdef} \.{VFDEBUG}\7
${}\\{LogError}(\.{EC\_MESS},\39\.{VFN},\39\.{"<voter\ \%d>\ :\ User\ m}\)%
\.{essage\ management:\ l}\)\.{oop\ \%d,\ code==\%d\\n"},\39\\{this\_voter},\39%
\|i,\39\\{msg}[\|i].\\{code});{}$\6
\8\#\&{endif}\6
\&{switch} ${}(\\{msg}[\|i].\\{code})$ $\{$ \6
\4\&{case} \.{VF\_INP\_MSG}:\6
\8\#\&{ifdef} \.{VFDEBUG}\6
${}\\{LogError}(\.{EC\_MESS},\39\.{VFN},\39\.{"<voter\ \%d>\ VF\_INP\_M}\)\.{SG%
\ received.\\n"},\39\\{this\_node});{}$\6
\8\#\&{endif}\6
${}\\{voter\_inputs}[\\{this\_voter}]\K\\{memdup}(\\{msg}[\|i].\\{msg},\39%
\\{msg}[\|i].\\{msglen});{}$\6
${}\\{input\_length}\K\\{msg}[\|i].\\{msglen};{}$\6
${}\\{input\_nr}\PP{}$;\C{ t0 = TimeNow(); }\6
\8\#\&{ifdef} \.{VFDEBUG}\6
${}\\{printf}(\.{"<voter\ \%d>\ message\ }\)\.{(as\ a\ double)\ is\ \%lf}\)\.{.%
\\n"},\39\\{this\_node},\39{*}{}$((\&{double} ${}{*}){}$ \\{msg}[\|i]${}.%
\\{msg}));{}$\6
\8\#\&{endif}\7
\X48:Check for a complete message suite; if so, vote, and possibly deliver the
outcome\X\7
${}\\{vf}\MG\\{inp\_msg\_got}\K\.{YES};{}$\6
\&{if} ${}(\\{this\_voter}\E\\{input\_nr}-\T{1}){}$\5
${}\{{}$\1\6
\&{int} \|j;\7
\X40:Broadcast the Input Message\X\6
${}\\{vf}\MG\\{broadcast\_done}\K\.{YES};{}$\6
\4${}\}{}$\2\6
\&{break};\7
\X41:case \PB{\.{VF\_DESTROY}}:\X\C{ \PB{\&{break};} unneeded, because the
statement is unreachable }\7
\X42:case \PB{\.{VF\_OUT\_LCB}}:\X \&{break};\7
\4\&{case} \.{VF\_SELECT\_ALG}:\5
${}\\{Algorithm}\K\\{msg}[\|i].\\{msglen};{}$\6
\&{break};\7
\4\&{case} \.{VF\_SCALING\_FACTOR}:\5
${}\\{ScalingFactor}\K{*}{}$((\&{double} ${}{*}){}$ \\{msg}[\|i]${}.%
\\{msg});{}$\6
\&{break};\7
\4\&{case} \.{VF\_EPSILON}:\5
${}\Epsilon\K{*}{}$((\&{double} ${}{*}){}$ \\{msg}[\|i]${}.\\{msg});{}$\6
\&{break};\7
\4\&{case} \.{VF\_RESET}:\5
${}\\{input\_nr}\K\\{input\_length}\K\T{0};{}$\6
${}\\{OutputLink}\K\NULL,\39\\{rvote}.\\{outcome}\K{*}\\{vote}\K\.{'\\0'};{}$\6
${}\\{vf}\MG\\{broadcast\_done}\K\\{vf}\MG\\{inp\_msg\_got}\K\\{vf}\MG%
\\{destroy\_requested}\K\.{NO};{}$\6
\8\#\&{ifndef} \.{STATIC}\6
${}\{{}$\1\6
\\{AllocationClass}\1\1\6
\&{int} \|i;\2\2\6
\&{for} ${}(\|i\K\T{0};{}$ ${}\|i<\\{vf}\MG\|N;{}$ ${}\|i\PP){}$\5
${}\{{}$\1\6
\\{free}(\\{voter\_inputs}[\|i]);\6
${}\\{voter\_inputs}[\|i]\K\NULL;{}$\6
\4${}\}{}$\2\6
\4${}\}{}$\2\6
\8\#\&{endif}\6
\&{break};\7
\4\&{case} \.{VF\_NOP}:\C{ for the time being, nothing }\6
\&{break};\7
\4\&{default}:\5
${}\\{printf}(\.{"<voter\ \%d>\ :\ defaul}\)\.{t\ case\ in\ switch:\ co}\)%
\.{de==\%d\\n"},\39\\{this\_voter},\39\\{msg}[\|i].\\{code});{}$\6
\&{break}; $\}{}$\C{ end \PB{\&{switch}} }\6
$\}{}$\C{ end \PB{\&{for}} }\par
\U38.\fi

\M{40}The actual transmission of this voter's input message to
all other voters in the farm.  The buffer is built up so to tie
the code of the message and the message itself.

\Y\B\4\X40:Broadcast the Input Message\X${}\E{}$\6
${*}{}$((\&{int} ${}{*}){}$ \\{buffer})${}\K\.{VF\_V\_INP\_MSG};{}$\6
${}\\{memcpy}(\\{buffer}+\&{sizeof}(\&{int}),\39\\{voter\_inputs}[\\{this%
\_voter}],\39\\{input\_length});{}$\6
\8\#\&{ifdef} \.{SERVERNET}\6
${}\\{Set\_Phase}(\\{serverlink},\39\.{VFP\_BROADCASTING});{}$\6
\8\#\&{endif}\6
\8\#\&{ifdef} \.{ZEROPERM}\6
\&{for} ${}(\|j\K\T{0};{}$ ${}\|j<\|N;{}$ ${}\|j\PP){}$\5
${}\{{}$\1\6
\&{if} ${}(\|j\I\\{this\_voter}){}$\5
${}\{{}$\6
\8\#\&{ifdef} \.{VFDEBUG}\1\6
${}\\{LogError}(\.{EC\_MESS},\39\.{VFN},\39\.{"<voter\ \%d>\ broadcas}\)\.{ting%
\ (j==\%d)\\n"},\39\\{this\_voter},\39\|j);{}$\6
\8\#\&{endif}\6
\&{if} ${}(\\{input\_length}+\&{sizeof}(\&{int})\I\\{SendLink}(\\{links}[\|j],%
\39\\{buffer},\39\\{input\_length}+\&{sizeof}(\&{int}))){}$\5
${}\{{}$\1\6
${}\\{LogError}(\.{EC\_ERROR},\39\.{VFN},\39\.{"Cannot\ SendLink\ to\ }\)%
\.{voter\ \%d."},\39\|j);{}$\6
\&{return} \\{VF\_error}${}\K\.{E\_VF\_SENDLINK};{}$\6
\4${}\}{}$\2\6
\4${}\}{}$\2\6
\4${}\}{}$\2\6
\8\#\&{else}\6
\&{for} ${}(\|j\K\\{this\_voter}+\T{1};{}$ ${}\|j<\|N;{}$ ${}\|j\PP){}$\5
${}\{{}$\1\6
\&{if} ${}(\\{input\_length}+\&{sizeof}(\&{int})\I\\{SendLink}(\\{links}[\|j],%
\39\\{buffer},\39\\{input\_length}+\&{sizeof}(\&{int}))){}$\5
${}\{{}$\1\6
${}\\{LogError}(\.{EC\_ERROR},\39\.{VFN},\39\.{"Cannot\ SendLink\ to\ }\)%
\.{voter\ \%d."},\39\|j);{}$\6
\&{return} \\{VF\_error}${}\K\.{E\_VF\_SENDLINK};{}$\6
\4${}\}{}$\2\6
\4${}\}{}$\2\6
\&{for} ${}(\|j\K\T{0};{}$ ${}\|j<\\{this\_voter};{}$ ${}\|j\PP){}$\5
${}\{{}$\1\6
\&{if} ${}(\\{input\_length}+\&{sizeof}(\&{int})\I\\{SendLink}(\\{links}[\|j],%
\39\\{buffer},\39\\{input\_length}+\&{sizeof}(\&{int}))){}$\5
${}\{{}$\1\6
${}\\{LogError}(\.{EC\_ERROR},\39\.{VFN},\39\.{"Cannot\ SendLink\ to\ }\)%
\.{voter\ \%d."},\39\|j);{}$\6
\&{return} \\{VF\_error}${}\K\.{E\_VF\_SENDLINK};{}$\6
\4${}\}{}$\2\6
\4${}\}{}$\2\6
\8\#\&{endif}\par
\Us39\ET45.\fi

\M{41}Management of a user message of type \PB{\.{VF\_DESTROY}}.
\Y\B\4\X41:case \PB{\.{VF\_DESTROY}}:\X${}\E{}$\6
\&{case} \.{VF\_DESTROY}:\C{ <Broadcast a \PB{\.{VF\_V\_DESTROY}} event> ?? }\6
\&{if} ${}(\\{vf}\MG\\{broadcast\_done}\E\.{NO}\W\\{vf}\MG\|N\I\T{1}){}$\5
${}\{{}$\1\6
${}\\{voter\_sendcode}(\\{UserLink},\39\.{VF\_REFUSED});{}$\6
\&{break};\6
\4${}\}{}$\2\6
\&{else} $\{{}$\6
\8\#\&{ifdef} \.{SERVERNET}\6
$\\{Set\_Phase}(\\{serverlink},\39\.{VFP\_QUITTING});{}$\6
\8\#\&{endif}\6
${}\\{voter\_sendcode}(\\{UserLink},\39\.{VF\_QUIT});{}$\6
\8\#\&{ifdef} \.{SERVERNET}\6
$\{$ \&{int} \\{myident}${}\K\\{vf}\MG\\{vf\_ident\_stack}[\\{this\_voter}];$ %
\&{int} \&{error}  ;\6
\8\#\&{ifdef} \\{DoBreakServer}\6
\&{if} ( ( \&{error} $\K$ \\{BreakServer}(\\{serverlink}) ) $\I$ \T{0} ) %
\\{LogError} $(\.{EC\_ERROR},\39\.{VFN},\39\.{"BreakServer\ failed,}\)\.{\
error:\%d"},\39$ \&{error} )  ;\6
\8\#\&{endif}\C{ DoBreakServer }\6
\8\#\&{ifdef} \.{VFDEBUG}\6
${}\\{LogError}(\.{EC\_MESS},\39\.{VFN},\39\.{"<voter\ \%d>\ myident=}\)\.{\%d.%
\ Bye.\\n"},\39\\{myident},\39\\{this\_voter});{}$\6
\8\#\&{endif}\6
$\}{}$\6
\8\#\&{endif}\C{ SERVERNET }\6
\\{exit}(\T{0});\6
\&{return} \\{VF\_error}${}\K\T{0};$ $\}{}$\par
\U39.\fi

\M{42}Management of a user message of type \PB{\.{VF\_OUT\_LCB}}.
\Y\B\4\X42:case \PB{\.{VF\_OUT\_LCB}}:\X${}\E{}$\6
\&{case} \.{VF\_OUT\_LCB}: \\{OutputLink} $\K$ ( $\LinkCBxt$ $*$ ) $\\{msg}[%
\|i].\\{msg};{}$\6
\&{if} ${}(\\{OutputLink}\I\NULL){}$\5
${}\{{}$\1\6
\&{if} ${}(\\{rvote}.\\{outcome}\E\.{VF\_SUCCESS}){}$\5
${}\{{}$\1\6
\X50:Deliver the Outcome\X\6
\4${}\}{}$\2\6
\4${}\}{}$\2\6
\&{else}\5
${}\{{}$\1\6
${}\\{LogError}(\.{EC\_ERROR},\39\.{VFN},\39\.{"Invalid\ output\ link}\)\.{\
control\ block\ -\ can}\)\.{'t\ deliver."});{}$\6
\4${}\}{}$\C{  In the event of a \PB{\.{VF\_DESTROY}} message, the local voter
should inform   its fellow via the broadcasting of a \PB{\.{VF\_V\_DESTROY}}
event. <Broadcast a \PB{\.{VF\_V\_DESTROY}} event>= }\2\par
\U39.\fi

\M{43}If the received options is not option \# \PB{\|N} nor option \# \PB{%
\\{this\_voter}},
then a message is coming from a voter in the farm.
Note that this time messages cannot be buffered---only one message
at a time will be received. Moreover, messages come from a different
memory space---for this reason, the structure of the message can't be
that of a \PB{\&{VF\_msg\_t}} object. A different approach must be used:
an integer representing the code message should be directly attached to an
opaque area. The resulting buffer constitutes the message.

\Y\B\4\X43:A message from the cliqu\'e\X${}\E{}$\6
\&{if} ${}(\\{opt}\I\\{this\_voter}\W\\{opt}\I\|N){}$\5
${}\{{}$\1\6
\&{if} ${}((\\{recv}\K\\{RecvLink}(\\{links}[\\{opt}],\39\\{buffer},\39\.{VF%
\_MAX\_INPUT\_MSG}))<\T{0}){}$\5
${}\{{}$\1\6
${}\\{LogError}(\.{EC\_ERROR},\39\.{VFN},\39\.{"Can't\ RecvLink\ [A\ m}\)%
\.{essage\ from\ the\ cliq}\)\.{ue]"});{}$\6
${}\\{LogError}(\.{EC\_ERROR},\39\.{VFN},\39\.{"\\t(Sender\ is\ voter\ }\)\.{%
\%d,\ size\ of\ message\ }\)\.{is\ \%d.)"},\39\\{opt},\39\\{recv});{}$\6
\&{return} \\{VF\_error}${}\K\.{E\_VF\_RECVLINK};{}$\6
\4${}\}{}$\2\6
${}\\{msg}[\T{0}].\\{code}\K{*}{}$((\&{int} ${}{*}){}$ \\{buffer});\6
${}\\{msg}[\T{0}].\\{msg}\K\\{buffer}+\&{sizeof}(\&{int});{}$\6
${}\\{msg}[\T{0}].\\{msglen}\K\\{recv}-\&{sizeof}(\&{int});{}$\6
\X44:Cliqu\'e message management\X\6
\4${}\}{}$\2\par
\U37.\fi

\M{44}A new message has come from a voter in the cliqu\'e.
\Y\B\4\X44:Cliqu\'e message management\X${}\E{}$\6
\&{switch} ${}(\\{msg}[\T{0}].\\{code}){}$\5
${}\{{}$\1\6
\4\&{case} \.{VF\_V\_INP\_MSG}:\6
\8\#\&{ifdef} \.{VFDEBUG}\6
${}\\{printf}(\.{"voter\ \%d\ received\ t}\)\.{he\ input\ message\ fro}\)\.{m\
voter\ \%d:\ \%lf\\n"},\39\\{this\_voter},\39\\{opt},\39{*}{}$((\&{double}
${}{*}){}$ \\{msg}[\T{0}]${}.\\{msg}));{}$\6
\8\#\&{endif}\6
\&{if} ${}(\\{input\_length}\E\T{0}){}$\1\5
${}\\{input\_length}\K\\{msg}[\T{0}].\\{msglen};{}$\2\6
\&{else}\5
${}\{{}$\1\6
\&{if} ${}(\\{input\_length}\I\\{msg}[\T{0}].\\{msglen}){}$\5
${}\{{}$\1\6
${}\\{LogError}(\.{EC\_ERROR},\39\.{VFN},\39\.{"Wrong\ input\ size"});{}$\6
\&{return} \\{VF\_error}${}\K\.{E\_VF\_INPUT\_SIZE};{}$\6
\4${}\}{}$\2\6
\4${}\}{}$\2\6
\8\#\&{ifndef} \.{STATIC}\6
\&{if} ${}((\\{voter\_inputs}[\\{opt}]\K{}$(\&{void} ${}{*}){}$ \\{malloc}(%
\\{input\_length}))${}\E\T{0}){}$\5
${}\{{}$\1\6
${}\\{LogError}(\.{EC\_ERROR},\39\.{VFN},\39\.{"Memory\ Allocation\ E}\)%
\.{rror."});{}$\6
\&{return} \\{VF\_error}${}\K\.{E\_VF\_CANT\_ALLOC};{}$\6
\4${}\}{}$\2\6
\8\#\&{else}\6
${}\\{voter\_inputs}[\\{opt}]\K{}$(\&{void} ${}{*}){}$ ${}{\AND}\\{st\_voter%
\_inputs\_data}[\\{opt}][\T{0}];{}$\6
\8\#\&{endif}\6
${}\\{memcpy}(\\{voter\_inputs}[\\{opt}],\39\\{msg}[\T{0}].\\{msg},\39\\{input%
\_length});{}$\6
${}\\{input\_nr}\PP{}$;\7
\X48:Check for a complete message suite; if so, vote, and possibly deliver the
outcome\X\6
\X45:Check if it's your turn to broadcast; if so, do it, and take note of that%
\X\6
\&{break};\6
\4\&{case} \.{VF\_V\_DESTROY}:\C{ Is there a use for such a message? }\C{ for
the time being, no action }\6
\8\#\&{ifdef} \.{VFDEBUG}\6
${}\\{LogError}(\.{EC\_MESS},\39\.{VFN},\39\.{"voter\ \%d\ received\ a}\)\.{\
VF\_V\_DESTROY\ messag}\)\.{e\ from\ voter\ \%d\\n"},\39\\{this\_voter},\39%
\\{opt});{}$\6
\8\#\&{endif}\6
\&{break};\6
\4\&{case} \.{VF\_V\_ERROR}:\C{ The voter notifies an error condition }\C{ for
the time being, no action }\6
\&{break};\6
\4\&{case} \.{VF\_V\_RESET}:\C{ Is there a use for such a message? }\C{ for the
time being, no action }\6
\&{break};\6
\4\&{case} \.{VF\_V\_NOP}:\C{ for the time being, no action }\6
\&{break};\6
\4${}\}{}$\C{ end \PB{\&{switch}} }\2\par
\U43.\fi

\M{45}Broadcasting is performed in an ordered fashion so to prevent
deadlocks---a
voter is allowed to broadcast only when the following two conditions hold at
once:
\item{$\bullet$} it has not performed a broadcast before, and
\item{$\bullet$} the vote-id (a number from 0 to \PB{$\\{vf}\MG\|N$}-1) is less
than or equal
to the current number of input messages that have been received (user message
included), minus one.
\Y\B\4\X45:Check if it's your turn to broadcast; if so, do it, and take note of
that\X${}\E{}$\C{ if (this\_voter <= input\_nr -1) }\6
\&{if} ${}(\\{this\_voter}\E\\{input\_nr}-\T{1}){}$\5
${}\{{}$\1\6
\&{if} ${}(\\{vf}\MG\\{inp\_msg\_got}\E\.{YES}\W\\{vf}\MG\\{broadcast\_done}\E%
\.{NO}){}$\5
${}\{{}$\1\6
\&{int} \|j;\7
\X40:Broadcast the Input Message\X\6
${}\\{vf}\MG\\{broadcast\_done}\K\.{YES};{}$\6
\4${}\}{}$\2\6
\4${}\}{}$\2\par
\U44.\fi

\M{46}If the received options is option \# \PB{\|N}, then a message is
coming from the local server module.

\Y\B\4\X46:A message from the server module\X${}\E{}$\6
\&{if} ${}(\\{opt}\E\|N){}$\5
${}\{{}$\1\6
\&{int} \\{recv};\6
\&{char} \\{msg}[\.{FTB\_ELEMENT\_SIZE}];\7
\&{if} ${}((\\{recv}\K\\{RecvLink}(\\{serverlink},\39{}$(\&{void} ${}{*}){}$ %
\\{msg}${},\39\.{FTB\_ELEMENT\_SIZE}))\Z\T{0}){}$\1\5
${}\\{LogError}(\.{EC\_ERROR},\39\.{VFN},\39\.{"couldn't\ receive\ a\ }\)%
\.{message\ from\ the\ ser}\)\.{ver\\n"});{}$\2\6
\&{else}\1\5
${}\\{LogError}(\.{EC\_DEBUG},\39\.{VFN},\39\.{"got\ server\ message\ }\)\.{of\
\%d\ bytes.\\n"},\39\\{recv});{}$\2\6
\X47:Server message management\X\6
\4${}\}{}$\2\par
\U37.\fi

\M{47}So far, an empty section.
\Y\B\4\X47:Server message management\X${}\E{}$\C{ should deal with message kept
in msg[], size recv. }\par
\U46.\fi

\M{48}Every time a new message has come and consequently \PB{\\{input\_nr}} has
been incremented,
we must check if it's time for voting and if so, after voting, we check if we
have an output address
\Y\B\4\X48:Check for a complete message suite; if so, vote, and possibly
deliver the outcome\X${}\E{}$\6
\8\#\&{ifdef} \.{VFDEBUG}\6
$\\{LogError}(\.{EC\_MESS},\39\.{VFN},\39\.{"<voter\ \%d>\ input\_nr}\)\.{==%
\%d,\ N==\%d\\n"},\39\\{this\_voter},\39\\{input\_nr},\39\|N);{}$\6
\8\#\&{endif}\6
\&{if} ${}(\\{input\_nr}\E\|N){}$\5
${}\{{}$\1\6
\X49:Perform Voting\X\6
\&{if} ${}(\\{OutputLink}\I\NULL){}$\5
${}\{{}$\1\6
\X50:Deliver the Outcome\X\6
\4${}\}{}$\C{ input\_nr = 0; }\2\6
\4${}\}{}$\2\par
\Us39\ET44.\fi

\M{49}The actual voting algorithm is managed by a function
whose address is kept in the \PB{\\{DoVoting}} array at entry
no. \PB{\\{Algorithm}}.
\Y\B\4\X49:Perform Voting\X${}\E{}$\6
\&{if} ${}(\\{Algorithm}\G\T{0}\W\\{Algorithm}<\.{VF\_NB\_ALGS}){}$\5
${}\{{}$\6
\8\#\&{ifdef} \.{SERVERNET}\1\6
${}\\{Set\_Phase}(\\{serverlink},\39\.{VFP\_VOTING});{}$\6
\8\#\&{endif}\6
${}\\{DoVoting}[\\{Algorithm}](\\{vf},\39{}$(\&{void} ${}{*}{*}){}$ \\{voter%
\_inputs}${},\39\\{input\_length},\39{\AND}\\{rvote});{}$\6
\4${}\}{}$\2\6
\&{else}\5
${}\{{}$\1\6
${}\\{LogError}(\.{EC\_ERROR},\39\.{VFN},\39\.{"Wrong\ Algorithm\ num}\)\.{ber:%
\ \%d,\ not\ in\ [0,\%}\)\.{d["},\39\\{Algorithm},\39\.{VF\_NB\_ALGS});{}$\6
\&{return} \\{VF\_error}${}\K\.{E\_VF\_WRONG\_ALGID};{}$\6
\4${}\}{}$\2\6
\8\#\&{ifdef} \.{VFDEBUG}\6
\&{if} ${}(\\{rvote}.\\{outcome}\E\.{VF\_SUCCESS}){}$\5
${}\{{}$\1\6
${}\\{LogError}(\.{EC\_MESS},\39\.{VFN},\39\.{"<voter\ \%d>\ is\ sendi}\)\.{ng\
a\ VF\_DONE\ msg\ to\ }\)\.{the\ user---vote\ ==\ \%}\)\.{lf.\\n"},\39\\{this%
\_voter},\39{*}{}$(\&{double} ${}{*}){}$ \\{rvote}${}.\\{vote});{}$\6
${}\\{printf}(\.{"vote==\%lf\\n"},\39{*}{}$(\&{double} ${}{*}){}$ \\{rvote}${}.%
\\{vote});{}$\6
\4${}\}{}$\2\6
\&{else}\1\5
${}\\{LogError}(\.{EC\_MESS},\39\.{VFN},\39\.{"Vote\ is\ undefined."});{}$\2\6
\8\#\&{endif}\6
${}\\{done}.\\{code}\K\.{VF\_DONE}{}$;\C{ possibly \PB{$\NULL$}, which means:
``no unique vote is available'' }\6
${}\\{done}.\\{msglen}\K\\{rvote}.\\{outcome};{}$\6
${}\\{done}.\\{msg}\K\\{rvote}.\\{vote};{}$\6
${}\\{recv}\K\\{voter\_sendmsg}(\\{UserLink},\39{\AND}\\{done});{}$\6
\8\#\&{ifdef} \.{VFDEBUG}\6
${}\\{LogError}(\.{EC\_MESS},\39\.{VFN},\39\.{"<voter\ \%d>\ sent\ a\ V}\)\.{F%
\_DONE\ msg\ (\%d\ bytes}\)\.{)\ to\ the\ user.\\n"},\39\\{this\_voter},\39%
\\{recv});{}$\6
\8\#\&{endif}\par
\U48.\fi

\M{50}The vote is sent to the output module.
\Y\B\4\X50:Deliver the Outcome\X${}\E{}$\6
\&{if} ${}(\\{rvote}.\\{outcome}\E\.{VF\_SUCCESS}){}$\5
${}\{{}$\1\6
\&{if} ${}(\\{input\_length}\I\\{SendLink}(\\{OutputLink},\39\\{vote},\39%
\\{input\_length})){}$\5
${}\{{}$\1\6
${}\\{LogError}(\.{EC\_ERROR},\39\.{VFN},\39\.{"Cannot\ deliver\ the\ }\)%
\.{output."});{}$\6
\&{return} \\{VF\_error}${}\K\.{E\_VF\_DELIVER};{}$\6
\4${}\}{}$\2\6
\8\#\&{ifdef} \.{SERVERNET}\6
${}\\{Set\_Phase}(\\{serverlink},\39\.{VFP\_WAITING});{}$\6
\8\#\&{endif}\6
\4${}\}{}$\2\6
\&{else}\5
${}\{{}$\1\6
\&{char} \|c${}\K\T{0};{}$\7
\&{if} ${}(\\{SendLink}(\\{OutputLink},\39{\AND}\|c,\39\T{1})\I\T{1}){}$\5
${}\{{}$\1\6
${}\\{LogError}(\.{EC\_ERROR},\39\.{VFN},\39\.{"Cannot\ deliver\ the\ }\)%
\.{negative\ result."});{}$\6
\&{return} \\{VF\_error}${}\K\.{E\_VF\_DELIVER};{}$\6
\4${}\}{}$\2\6
\8\#\&{ifdef} \.{SERVERNET}\6
${}\\{Set\_Phase}(\\{serverlink},\39\.{VFP\_FAILED});{}$\6
\8\#\&{endif}\6
\4${}\}{}$\2\par
\Us42\ET48.\fi

\M{51}\PB{\\{VF\_perror}}: the \PB{\\{perror}} function of the VotingFarm
class.
Statically defined as a vector of strings, its entries can be addressed
as ``$-e$'', where $e$ is the error condition returned in \PB{\\{VF\_error}}.
The number of messages has been specified so to reduce the risk of
inconsistencies.

\Y\B\4\X51:Voting Farm Error Function\X${}\E{}$\6
\&{static} \&{char} ${}{*}\\{errors}[\.{VF\_ERROR\_NB}]\K\{\.{"no\ error"},{}$%
\C{ no error }\6
\.{"An\ internal\ stack\ h}\)\.{as\ reached\ its\ upper}\)\.{\ limit"}${},{}$%
\C{ \PB{\.{E\_VF\_OVERFLOW}} }\6
\.{"The\ system\ was\ not\ }\)\.{able\ to\ execute\ allo}\)\.{cation"}${},{}$%
\C{ \PB{\.{E\_VF\_CANT\_ALLOC}} }\6
\.{"This\ operation\ requ}\)\.{ires\ a\ defined\ votin}\)\.{g\ farm"}${},{}$\C{
\PB{\.{E\_VF\_UNDEFINED\_VF}} }\6
\.{"A\ wrong\ node\ has\ be}\)\.{en\ specified"}${},{}$\C{ \PB{\.{E\_VF\_WRONG%
\_NODE}} }\6
\.{"The\ system\ was\ not\ }\)\.{able\ to\ get\ the\ glob}\)\.{al\ id"}${},{}$%
\C{ \PB{\.{E\_VF\_GETGLOBID}} }\6
\.{"The\ system\ was\ not\ }\)\.{able\ to\ execute\ Crea}\)\.{teThread"}${},{}$%
\C{ \PB{\.{E\_VF\_CANT\_SPAWN}} }\6
\.{"The\ system\ was\ not\ }\)\.{able\ to\ execute\ Conn}\)\.{ectLink"}${},{}$%
\C{ \PB{\\{ConnectLink}} error }\6
\.{"The\ system\ was\ not\ }\)\.{able\ to\ execute\ Recv}\)\.{Link"}${},{}$\C{ %
\PB{\.{E\_VF\_RECVLINK}} }\6
\.{"The\ system\ was\ not\ }\)\.{able\ to\ perform\ broa}\)\.{dcasting"}${},{}$%
\C{ \PB{\.{E\_VF\_BROADCAST}} }\6
\.{"Invalid\ output\ (Lin}\)\.{kCB\_t*)\ -\ can't\ deli}\)\.{ver"}${},{}$\C{ %
\PB{\.{E\_VF\_DELIVER}} }\6
\.{"Duplicated\ input\ me}\)\.{ssage"}${},{}$\C{ \PB{\.{E\_VF\_BUSY\_SLOT}} }\6
\.{"Invalid\ voting\ farm}\)\.{\ id"}${},{}$\C{ \PB{\.{E\_VF\_WRONG\_VFID}} }\6
\.{"Invalid\ metric\ func}\)\.{tion\ pointer"}${},{}$\C{ \PB{\.{E\_VF\_WRONG%
\_DISTANCE}} }\6
\.{"Inconsistent\ voting}\)\.{\ farm\ object"}${},{}$\C{ \PB{\.{E\_VF\_INVALID%
\_VF}} }\6
\.{"No\ local\ voters---o}\)\.{ne\ voter\ has\ to\ be\ s}\)\.{pecified"}${},{}$%
\C{ \PB{\.{E\_VF\_NO\_LVOTER}} }\6
\.{"More\ than\ one\ local}\)\.{\ voter\ has\ been\ spec}\)\.{ified"}${},{}$\C{
\PB{\.{E\_VF\_TOO\_MANY\_LVOTER}} }\6
\.{"A\ wrong\ number\ of\ m}\)\.{essages\ has\ been\ spe}\)\.{cified"}${},{}$%
\C{ \PB{\.{E\_VF\_WRONG\_MSG\_NB}} }\6
\.{"The\ system\ was\ not\ }\)\.{able\ to\ execute\ Send}\)\.{Link"}${},{}$\C{ %
\PB{\.{E\_VF\_SENDLINK}} }\6
\.{"Inconsistency\ in\ th}\)\.{e\ size\ of\ the\ input\ }\)\.{message"}${},{}$%
\C{ \PB{\.{E\_VF\_INPUT\_SIZE}} }\6
\.{"This\ operation\ requ}\)\.{ires\ a\ described\ vot}\)\.{ing\ farm"}${},{}$%
\C{ \PB{\.{E\_VF\_UNDESCRIBED}} }\6
\.{"This\ operation\ requ}\)\.{ires\ an\ active\ votin}\)\.{g\ farm"}${},{}$\C{
\PB{\.{E\_VF\_INACTIVE}} }\6
\.{"Inconsistency\ -\ sen}\)\.{der\ unknown"}${},{}$\C{ \PB{\.{E\_VF\_UNKNOWN%
\_SENDER}} }\6
\.{"Time-out\ reached\ du}\)\.{ring\ a\ Select()"}${},{}$\C{ \PB{\.{E\_VF%
\_EVENT\_TIMEOUT}} }\6
\.{"A\ Select()\ returned}\)\.{\ an\ index\ out\ of\ ran}\)\.{ge"}${},{}$\C{ %
\PB{\.{E\_VF\_SELECT}} }\6
\.{"Algorithm\ Id\ out\ of}\)\.{\ range"}${},{}$\C{ \PB{\.{E\_VF\_WRONG%
\_ALGID}} }\6
\.{"NULL\ in\ a\ call-by-r}\)\.{eference\ pointer"}${},{}$\C{ \PB{\.{E\_VF%
\_NULLPTR}} }\6
\.{"Maximun\ number\ of\ o}\)\.{pened\ voting\ farms\ e}\)\.{xceeded"}${},{}$%
\C{ \PB{\.{E\_VF\_TOO\_MANY}} }\6
${}\};{}$\7
\&{void} \\{VF\_perror}(\&{void})\1\1\2\2\6
${}\{{}$\1\6
\&{static} \&{char} ${}{*}\.{VFN}\K\.{"VF\_perror"};{}$\7
\&{if} (\\{VF\_error})\5
${}\{{}$\1\6
${}\\{fprintf}(\\{stderr},\39\.{"Error\ condition\ num}\)\.{ber\ \%d\ raised\
while\ }\)\.{in\ function\ \%s:\ \\"\%s}\)\.{\\"\\n"},\39\\{VF\_error},\39%
\.{VFN},\39\\{errors}[{-}\\{VF\_error}]);{}$\6
\\{fflush}(\\{stderr});\6
\4${}\}{}$\2\6
\4${}\}{}$\2\par
\U1.\fi

\M{52}The functions stored in the \PB{\\{DoVoting}} array are defined here.
\Y\B\4\X52:Voting Functions\X${}\E{}$\6
\X54:Exact Concensus\X\6
\X55:Majority Voting\X\6
\X56:Median Voting\X\6
\X57:Plurality Voting\X\6
\X58:Weighted Averaging\X\6
\X53:Simple Majority Voting\X\6
\X60:Simple Average\X\par
\U29.\fi

\M{53}The simplest algorithm, apart from exact concensus---counts
the agreement and returns the widest.
\Y\B\4\X53:Simple Majority Voting\X${}\E{}$\6
\&{static} \&{void} \\{VFA\_SimpleMajorityVoting}(\&{VotingFarm\_t} ${}{*}%
\\{vf},\39{}$\&{void} ${}{*}\\{inp}[\,],\39{}$\&{int} \\{len}${},\39{}$\&{vote%
\_t} ${}{*}\\{vote}){}$\1\1\2\2\6
${}\{{}$\1\6
\&{int} \|i${},{}$ \|j;\6
\&{int} \|n${}\K\\{vf}\MG\|N;{}$\6
\&{int} \|v[\.{VF\_MAX\_NTS}];\6
\&{int} \\{threshold};\7
${}\\{threshold}\K\|n\GG\T{1}{}$;\C{ n/2; }\6
\&{for} ${}(\|i\K\T{0};{}$ ${}\|i<\|n;{}$ ${}\|i\PP){}$\1\5
${}\|v[\|i]\K\T{0};{}$\2\6
\&{for} ${}(\|i\K\T{0};{}$ ${}\|i<\|n;{}$ ${}\|i\PP){}$\1\6
\&{for} ${}(\|j\K\T{0};{}$ ${}\|j<\|n;{}$ ${}\|j\PP){}$\1\6
\&{if} ${}(\|i\I\|j){}$\5
${}\{{}$\1\6
\&{if} ${}(\\{vf}\MG\\{distance}(\\{inp}[\|i],\39\\{inp}[\|j])<\.{VFD%
\_EPSILON}){}$\5
${}\{{}$\1\6
${}\|v[\|i]\PP;{}$\6
\4${}\}{}$\2\6
\4${}\}{}$\2\2\2\6
\&{for} ${}(\|i\K\T{0};{}$ ${}\|i<\|n;{}$ ${}\|i\PP){}$\1\6
\&{if} ${}(\|v[\|i]\G\\{threshold}){}$\5
${}\{{}$\1\6
${}\\{vote}\MG\\{outcome}\K\.{VF\_SUCCESS};{}$\6
${}\\{memcpy}(\\{vote}\MG\\{vote},\39\\{inp}[\|i],\39\\{len});{}$\6
\&{return};\6
\4${}\}{}$\2\2\6
${}\\{vote}\MG\\{outcome}\K\.{VF\_FAILURE};{}$\6
\4${}\}{}$\2\par
\U52.\fi

\M{54}Exact concensus means perfect, bitwise equality.
\Y\B\4\X54:Exact Concensus\X${}\E{}$\6
\&{static} \&{void} \\{VFA\_ExactConcensus}(\&{VotingFarm\_t} ${}{*}\\{vf},%
\39{}$\&{void} ${}{*}\\{inp}[\,],\39{}$\&{int} \\{len}${},\39{}$\&{vote\_t}
${}{*}\\{vote}){}$\1\1\2\2\6
${}\{{}$\1\6
\&{int} \|i;\6
\&{int} \|n${}\K\\{vf}\MG\|N;{}$\7
\&{if} ${}(\\{inp}[\T{0}]\E\NULL){}$\5
${}\{{}$\1\6
${}\\{vote}\MG\\{outcome}\K\.{VF\_FAILURE};{}$\6
\&{return};\6
\4${}\}{}$\2\6
\&{for} ${}(\|i\K\T{1};{}$ ${}\|i<\|n;{}$ ${}\|i\PP){}$\5
${}\{{}$\1\6
\&{if} ${}(\\{inp}[\|i]\E\NULL){}$\5
${}\{{}$\1\6
${}\\{vote}\MG\\{outcome}\K\.{VF\_FAILURE};{}$\6
\&{return};\6
\4${}\}{}$\2\6
\&{if} ${}(\\{memcmp}(\\{inp}[\T{0}],\39\\{inp}[\|i],\39\\{len})\I\T{0}){}$\5
${}\{{}$\1\6
${}\\{vote}\MG\\{outcome}\K\.{VF\_FAILURE};{}$\6
\&{return};\6
\4${}\}{}$\2\6
\4${}\}{}$\2\6
${}\\{vote}\MG\\{outcome}\K\.{VF\_SUCCESS};{}$\6
${}\\{memcpy}(\\{vote}\MG\\{vote},\39\\{inp}[\T{0}],\39\\{len});{}$\6
\4${}\}{}$\2\par
\U52.\fi

\N{1}{55}Generalized Voters.
Several commonly used voting techniques have been generalized in [{\sc 10}]
to ``arbitrary $N$-version systems with arbitrary output types using a
metric space framework'', including:

\item{$\bullet$} formalized majority voter (\PB{\.{VFA\_MAJORITY}};
cf. [{\sc 10}, \S2.1, pp.445--446]),
\item{$\bullet$} generalized median voter (\PB{\.{VFA\_MEDIAN}};
cf. [{\sc 10}, \S2.2, p.447]),
\item{$\bullet$} formalized plurality voter (\PB{\.{VFA\_PLURALITY}};
cf. [{\sc 10}, \S2.3, pp.447--448]), and the
\item{$\bullet$} weighted averaging technique (\PB{\.{VFA\_WEIGHTED\_AVG}};
cf. [{\sc 10}, \S2.4, p.448]).

All these techniques are based on the concept of ``metric space'' which is
now recalled:

A metric space is a couple $(X,d)$, where $X$ is the output space
of the voting threads and $d$ is a real value function defined on $X\times X$
which is able in some way to ``compare'' two objects belonging to $X$;
more precisely, $d$ behaves as a ``distance'' measure of any two
objects in $X$. More formally,
$\forall (x,y,z)\in X^3$ the following properties hold:

\vskip 6pt

\item{1.} $d(x,y)\ge0$ (distances are positive numbers or zeroes);
\item{2.} $d(x,y)=0 \Rightarrow x=y$ (different points have positive
distances);
\item{3.} $d(x,y)=d(y,x)$ (distances obey the reflexive property);
\item{4.} $d(x,z)\le d(x,y)+d(y,z)$ (two consecutive segments are greater
than the segment that straightly connects their loose ends,
unless the three points lie on the same straight line;)

\vskip 6pt

\noindent
then $d$ is called a ``metric''.

In other words, a metric is a function which is able to compare any two input
objects and is able to numerically express a degree of ``closeness'' between
them.
[{\sc 10}] shows how four different voting algorithms can be executed
starting
from such a function.
It is the user responsibility to supply a valid metric function
on the call to \PB{\\{VF\_open}}: that function
shall get two pointers to opaque objets,
compute a ``distance'', and return that value as a positive
real number.

These functions take advantage of the \PB{\\{Stack}} class which has been used
in order to mimic the list operations in the algorithms in [{\sc 10}].

\Y\B\4\X55:Majority Voting\X${}\E{}$\6
\&{static} \&{void} \\{VFA\_MajorityVoting}(\&{VotingFarm\_t} ${}{*}\\{vf},%
\39{}$\&{void} ${}{*}\\{inp}[\,],\39{}$\&{int} \\{len}${},\39{}$\&{vote\_t}
${}{*}\\{vote}){}$\1\1\2\2\6
${}\{{}$\1\6
\&{int} \|i;\6
\&{int} \|n${}\K\\{vf}\MG\|N;{}$\6
\&{int} \|v;\6
\&{cluster\_t} ${}{*}\|c;{}$\6
\8\#\&{ifndef} \.{STATIC}\7
${}\|c\K\\{calloc}(\|n,\39\&{sizeof}(\&{cluster\_t}));{}$\6
\8\#\&{else}\6
${}\|c\K\\{st\_clusters};{}$\6
${}\\{memset}(\|c,\39\T{0},\39\|n*\&{sizeof}(\&{cluster\_t}));{}$\6
\8\#\&{endif}\6
\X59:Create a partition of blocks which are maximal with respect to the metric
property\X\C{ v is set by $<$Create a partition...$>$ to the cardinality of the
partition }\6
\&{for} ${}(\|i\K\T{0};{}$ ${}\|i<\|v;{}$ ${}\|i\PP){}$\5
${}\{{}$\1\6
\&{if} ${}(\|c[\|i].\\{item\_nr}>\|n/\T{2}){}$\5
${}\{{}$\1\6
${}\\{vote}\MG\\{outcome}\K\.{VF\_SUCCESS};{}$\6
${}\\{memcpy}(\\{vote}\MG\\{vote},\39\|c[\|i].\\{item},\39\\{len});{}$\6
\&{return};\6
\4${}\}{}$\2\6
\4${}\}{}$\2\6
${}\\{vote}\MG\\{outcome}\K\.{VF\_FAILURE};{}$\6
\4${}\}{}$\2\par
\U52.\fi

\M{56}``Generalized Median Voter'', \S2.2 of [{\sc 10}, p.447].
See also the ``mid-value select'' technique in [{\sc 9}, p.60]
\Y\B\4\D$\.{PRESENT}$ \5
\T{1}\par
\B\4\D$\.{NOT\_PRESENT}$ \5
\T{0}\par
\Y\B\4\X56:Median Voting\X${}\E{}$\6
\&{static} \&{void} \\{VFA\_MedianVoting}(\&{VotingFarm\_t} ${}{*}\\{vf},\39{}$%
\&{void} ${}{*}\\{inp}[\,],\39{}$\&{int} \\{len}${},\39{}$\&{vote\_t} ${}{*}%
\\{vote}){}$\1\1\2\2\6
${}\{{}$\1\6
\&{void} ${}{*}\\{inputs}[\.{VF\_MAX\_NTS}];{}$\6
\&{value\_t} ${}{*}\|v;{}$\6
\&{int} \|i${},{}$ \|j;\6
\&{int} \\{ri}${},{}$ \\{rj};\6
\&{double} \\{max}${},{}$ \\{dist};\6
\&{int} \|n${},{}$ \\{card};\6
\8\#\&{ifndef} \.{STATIC}\6
\&{static} \&{char} ${}{*}\.{VFN}\K\.{"VFA/MedianVoting"};{}$\6
\8\#\&{endif}\6
\8\#\&{ifdef} \.{STATIC}\7
${}\|v\K\\{st\_VFA\_v};{}$\6
\8\#\&{else}\6
${}\|v\K{}$(\&{value\_t} ${}{*}){}$ \\{malloc}${}(\\{len}*\&{sizeof}(\&{value%
\_t}));{}$\6
\&{if} ${}(\|v\E\NULL){}$\5
${}\{{}$\1\6
${}\\{LogError}(\.{EC\_ERROR},\39\.{VFN},\39\.{"Memory\ Allocation\ E}\)%
\.{rror."});{}$\6
${}\\{VF\_error}\K\.{E\_VF\_CANT\_ALLOC};{}$\6
\&{return};\6
\4${}\}{}$\2\6
\8\#\&{endif}\C{ \PB{\.{STATIC}} }\6
\&{for} ${}(\|n\K\\{vf}\MG\|N,\39\|i\K\T{0};{}$ ${}\|i<\|n;{}$ ${}\|i\PP){}$\5
${}\{{}$\1\6
${}\|v[\|i].\\{object}\K\\{inp}[\|i];{}$\6
${}\|v[\|i].\\{status}\K\.{PRESENT};{}$\6
\4${}\}{}$\2\6
${}\\{ri}\K\\{rj}\K\T{0};{}$\6
\&{do}\5
${}\{{}$\1\6
\&{for} ${}(\\{card}\K\|i\K\T{0},\39\\{max}\K{-}\T{1.0};{}$ ${}\|i<\|n;{}$ ${}%
\|i\PP){}$\5
${}\{{}$\1\6
\&{if} ${}(\|v[\|i].\\{status}\E\.{PRESENT}){}$\5
${}\{{}$\1\6
${}\\{inputs}[\\{card}\PP]\K\|v[\|i].\\{object};{}$\6
\&{for} ${}(\|j\K\|i+\T{1};{}$ ${}\|j<\|n;{}$ ${}\|j\PP){}$\5
${}\{{}$\1\6
\&{if} ${}(\|v[\|j].\\{status}\E\.{PRESENT}){}$\1\6
\&{if} ${}((\\{dist}\K(\\{vf}\MG\\{distance})(\|v[\|i].\\{object},\39\|v[\|j].%
\\{object}))\G\\{max}){}$\5
${}\{{}$\1\6
${}\\{max}\K\\{dist};{}$\6
${}\\{ri}\K\|i;{}$\6
${}\\{rj}\K\|j;{}$\6
\4${}\}{}$\2\2\6
\4${}\}{}$\2\6
\4${}\}{}$\2\6
\4${}\}{}$\2\6
\&{if} ${}(\\{max}\I{-}\T{1.0}){}$\5
${}\{{}$\1\6
${}\|v[\\{ri}].\\{status}\K\|v[\\{rj}].\\{status}\K\.{NOT\_PRESENT};{}$\6
\4${}\}{}$\2\6
\4${}\}{}$\2\5
\&{while} ${}(\\{card}>\T{2});{}$\6
${}\\{vote}\MG\\{outcome}\K\.{VF\_SUCCESS};{}$\6
${}\\{memcpy}(\\{vote}\MG\\{vote},\39\\{inputs}[\T{0}],\39\\{len});{}$\6
\4${}\}{}$\2\par
\U52.\fi

\M{57}``Formalized Plurality Voter'', \S2.3 of [{\sc 10}, p.447].
\Y\B\4\X57:Plurality Voting\X${}\E{}$\6
\&{static} \&{void} \\{VFA\_PluralityVoting}(\&{VotingFarm\_t} ${}{*}\\{vf},%
\39{}$\&{void} ${}{*}\\{inp}[\,],\39{}$\&{int} \\{len}${},\39{}$\&{vote\_t}
${}{*}\\{vote}){}$\1\1\2\2\6
${}\{{}$\1\6
\&{int} \|i${},{}$ \|j;\6
\&{int} \|n${}\K\\{vf}\MG\|N;{}$\6
\&{int} \|v;\6
\&{int} \\{max};\6
\&{cluster\_t} ${}{*}\|c;{}$\6
\8\#\&{ifndef} \.{STATIC}\7
${}\|c\K\\{calloc}(\|n,\39\&{sizeof}(\&{cluster\_t}));{}$\6
\8\#\&{else}\6
${}\|c\K\\{st\_clusters};{}$\6
${}\\{memset}(\|c,\39\T{0},\39\|n*\&{sizeof}(\&{cluster\_t}));{}$\6
\8\#\&{endif}\6
\&{if} ${}(\|c\E\NULL){}$\1\5
${}\\{LogError}(\.{EC\_MESS},\39\.{"Plurality"},\39\.{"c\ is\ NULL"});{}$\2\6
\X59:Create a partition of blocks which are maximal with respect to the metric
property\X\6
\8\#\&{ifdef} \.{VFDEBUG}\6
${}\\{LogError}(\.{EC\_MESS},\39\.{"Plurality"},\39\.{"Partition\ has\ been\ }%
\)\.{created."});{}$\6
\8\#\&{endif}\6
${}\|j\K{-}\T{1};{}$\6
\&{for} ${}(\\{max}\K\|i\K\T{0};{}$ ${}\|i<\|v;{}$ ${}\|i\PP){}$\5
${}\{{}$\1\6
\&{if} ${}(\|c[\|i].\\{item\_nr}>\\{max}){}$\1\5
${}\|j\K\|i,\39\\{max}\K\|c[\|i].\\{item\_nr};{}$\2\6
\4${}\}{}$\2\6
\8\#\&{ifdef} \.{VFDEBUG}\6
${}\\{LogError}(\.{EC\_MESS},\39\.{"Plurality"},\39\.{"Max\ computed."});{}$\6
\8\#\&{endif}\6
\&{if} ${}(\\{max}>\T{1}){}$\5
${}\{{}$\1\6
${}\\{vote}\MG\\{outcome}\K\.{VF\_SUCCESS}{}$;\C{ memcpy(vote->vote, c[i].item,
len); }\6
${}\\{memcpy}(\\{vote}\MG\\{vote},\39\|c[\|j].\\{item},\39\\{len});{}$\6
\4${}\}{}$\2\6
\&{else}\1\5
${}\\{vote}\MG\\{outcome}\K\.{VF\_FAILURE};{}$\2\6
\8\#\&{ifndef} \.{STATIC}\6
\\{free}(\|c);\6
\8\#\&{endif}\6
\4${}\}{}$\2\par
\U52.\fi

\M{58}``Weighted Averaging Technique'', \S2.4 of [{\sc 10}, p.448].
The \PB{\\{ScalingFactor}} variable is used for computing a set of ``weights'',
defined as follows:
given $n$ values, $x_1, x_2, \dots, x_n$, then
$$\forall i\in\{1,2,\dots,n\}:
w_i = \big[ 1+{\prod_{j=1,j\neq i}^n {\hbox{\bf d}^2(x_j, x_i)}\over{a}} %
\big]^{-1}$$
where $a$ is equal to \PB{\\{ScalingFactor}} and {\bf d} is
the metric. Considered $S=\sum_{i=1}^n w_i$, the voted value is computed
as $x = \big({\sum_{i=1}^n{w_i}\over{S}}\big)x_i$ which is of course here
computed
as ${{\sum_{i=1}^n w_i x_i}\over{S}}$.

\Y\B\4\X58:Weighted Averaging\X${}\E{}$\6
\&{static} \&{void} \\{VFA\_WeightedAveraging}(\&{VotingFarm\_t} ${}{*}\\{vf},%
\39{}$\&{void} ${}{*}\\{inp}[\,],\39{}$\&{int} \\{len}${},\39{}$\&{vote\_t}
${}{*}\\{vote}){}$\1\1\2\2\6
${}\{{}$\1\6
\&{int} \|i${},{}$ \|j;\6
\&{int} \|n${}\K\\{vf}\MG\|N;{}$\6
\&{double} ${}{*}\\{sum};{}$\6
\&{double} ${}{*}\\{weight},{}$ \\{wsum};\6
\&{double} ${}{*}\\{squaredist};{}$\6
\&{double} \\{partial}${},{}$ \|f;\6
\&{int} \|r${},{}$ \|c;\6
\&{static} \&{char} ${}{*}\.{VFN}\K\.{"WeightedAveraging"};{}$\6
\8\#\&{ifdef} \.{STATIC}\7
${}\\{sum}\K{\AND}\\{st\_VFA\_sum};{}$\6
${}\\{weight}\K\\{st\_VFA\_weight};{}$\6
${}\\{squaredist}\K\\{st\_VFA\_squaredist};{}$\6
\8\#\&{else}\6
${}\\{sum}\K\\{malloc}(\&{sizeof}(\&{double}));{}$\6
${}\\{weight}\K{}$(\&{double} ${}{*}){}$ \\{malloc}${}(\|n*\&{sizeof}(%
\&{double}));{}$\6
${}\\{squaredist}\K{}$(\&{double} ${}{*}){}$ \\{malloc}${}(\|n*\|n*\&{sizeof}(%
\&{double}));{}$\6
\8\#\&{endif}\C{ \PB{\.{STATIC}} }\6
\&{if} ${}(\\{sum}\E\NULL\V\\{weight}\E\NULL\V\\{squaredist}\E\NULL){}$\5
${}\{{}$\1\6
${}\\{LogError}(\.{EC\_ERROR},\39\.{VFN},\39\.{"Memory\ Allocation\ E}\)%
\.{rror."});{}$\6
${}\\{VF\_error}\K\.{E\_VF\_CANT\_ALLOC};{}$\6
\&{return};\6
\4${}\}{}$\2\6
\&{if} ${}(\\{ScalingFactor}\E\T{0}){}$\5
${}\{{}$\1\6
${}\\{LogError}(\.{EC\_MESS},\39\.{VFN},\39\.{"Illegal\ scaling\ fac}\)\.{tor\
---\ set\ to\ 1"});{}$\6
${}\\{ScalingFactor}\K\T{1.0};{}$\6
\4${}\}{}$\C{ compute the distances }\2\6
\&{for} ${}(\|i\K\T{0};{}$ ${}\|i<\|n;{}$ ${}\|i\PP){}$\1\6
\&{for} ${}(\|j\K\T{0};{}$ ${}\|j<\|i;{}$ ${}\|j\PP){}$\5
${}\{{}$\1\6
${}\|f\K(\\{vf}\MG\\{distance})(\\{inp}[\|i],\39\\{inp}[\|j]);{}$\6
${}\\{squaredist}[\|i*\|n+\|j]\K\|f*\|f;{}$\6
\4${}\}{}$\2\2\6
\&{for} ${}(\\{wsum}\K\T{0.0},\39\|i\K\T{0};{}$ ${}\|i<\|n;{}$ ${}\|i\PP){}$\5
${}\{{}$\1\6
${}\\{partial}\K\T{1.0};{}$\6
\&{for} ${}(\|j\K\T{0};{}$ ${}\|j<\|n\W\|j\I\|i;{}$ ${}\|j\PP){}$\5
${}\{{}$\1\6
\&{if} ${}(\|i<\|j){}$\1\5
${}\|r\K\|j,\39\|c\K\|i;{}$\2\6
\&{else}\1\5
${}\|r\K\|i,\39\|c\K\|j;{}$\2\6
${}\\{partial}\MRL{*{\K}}\\{squaredist}[\|r*\|n+\|c];{}$\6
\4${}\}{}$\2\6
${}\\{partial}\MRL{{/}{\K}}(\\{ScalingFactor}*\\{ScalingFactor});{}$\6
${}\\{wsum}\MRL{+{\K}}\\{weight}[\|i]\K\T{1.0}/(\T{1.0}+\\{partial});{}$\6
\4${}\}{}$\2\6
\&{for} ${}({*}\\{sum}\K\T{0.0},\39\|i\K\T{0};{}$ ${}\|i<\|n;{}$ ${}\|i\PP){}$\5
${}\{{}$\1\6
${}{*}\\{sum}\MRL{+{\K}}({*}{}$(\&{double} ${}{*}){}$ \\{inp}[\|i])${}*%
\\{weight}[\|i];{}$\6
\4${}\}{}$\2\6
\&{if} ${}(\\{wsum}\I\T{0}){}$\5
${}\{{}$\1\6
${}{*}\\{sum}\MRL{{/}{\K}}\\{wsum};{}$\6
${}\\{vote}\MG\\{outcome}\K\.{VF\_SUCCESS};{}$\6
${}\\{memcpy}(\\{vote}\MG\\{vote},\39\\{sum},\39\\{len});{}$\6
\4${}\}{}$\2\6
\&{else}\1\5
${}\\{vote}\MG\\{outcome}\K\.{VF\_FAILURE};{}$\2\6
\8\#\&{ifndef} \.{STATIC}\6
\\{free}(\\{sum});\6
\\{free}(\\{weight});\6
\\{free}(\\{squaredist});\6
\8\#\&{endif}\6
\4${}\}{}$\2\par
\U52.\fi

\M{59}The input values are partitioned into a set of blocks, $V_1, V_2, \dots,
V_n$, such that
for each $i$ block $V_i$ is maximal with respect to the property that
$$\forall (x,y)\in V_i\times V_i : \hbox{\bf d}(x,y) \le \epsilon,$$
where {\bf d} is the metric.

In order to reproduce as much as possible the algorithmic formalism of [{\sc
10}] we decided
\item{$\bullet$} to mimic their Lisp-like statements with stacks, and
\item{$\bullet$} to use \PB{\&{goto}} statements.

In this way actions (1)--(6) in [{\sc 10}, p. 445--446] can be (more or
less)
mapped into the statements corresponding to labels \PB{\\{one}} to \PB{\\{six}}
that follow.

\Y\B\4\X59:Create a partition of blocks which are maximal with respect to the
metric property\X${}\E{}$\6
${}\{{}$\1\6
\&{char} \\{vt};\6
\&{char} ${}{*}\\{del};{}$\6
\&{void} ${}{*}\\{item};{}$\6
\&{int} \|i${},{}$ \|j${},{}$ \\{item\_nr};\7
${}\\{vt}\K\.{GET\_ROOT}(\,)\MG\\{ProcRoot}\MG\\{MyProcID};{}$\6
\8\#\&{ifndef} \.{STATIC}\6
${}\\{del}\K\\{calloc}(\|n,\39\T{1}){}$;\C{ alloc + set all of them to NO }\6
\8\#\&{else}\6
${}\\{del}\K\\{st\_chars};{}$\6
${}\\{memset}(\\{del},\39\T{0},\39\|n);{}$\6
\8\#\&{endif}\6
\&{for} ${}(\|v\K\|i\K\T{0};{}$ ${}\|i<\|n;{}$ ${}\|i\PP){}$\5
${}\{{}$\1\6
\&{if} (\\{del}[\|i])\1\5
\&{continue};\2\6
${}\\{item}\K\\{inp}[\|i];{}$\6
${}\\{del}[\|i]\K\.{YES};{}$\6
${}\|c[\|v].\\{item}\K\\{item};{}$\6
\&{for} ${}(\\{item\_nr}\K\T{1},\39\|j\K\|i+\T{1};{}$ ${}\|j<\|n;{}$ ${}\|j%
\PP){}$\5
${}\{{}$\1\6
\&{if} ${}(\R\\{del}[\|j]\W\\{vf}\MG\\{distance}(\\{item},\39\\{inp}[\|j])<%
\Epsilon){}$\5
${}\{{}$\1\6
${}\\{del}[\|j]\K\.{YES};{}$\6
${}\\{item\_nr}\PP;{}$\6
\4${}\}{}$\2\6
\4${}\}{}$\2\6
${}\|c[\|v].\\{item\_nr}\K\\{item\_nr};{}$\6
${}\|v\PP;{}$\6
\4${}\}{}$\2\6
\8\#\&{ifndef} \.{STATIC}\6
\\{free}(\\{del});\6
\8\#\&{endif}\6
\4${}\}{}$\2\par
\Us55\ET57.\fi

\M{60}Simple Averaging may be useful e.g., to ``melt together'' $n$ sample
values
of a same pixel of an image. Of course it requires the objects are numbers.
For the time being, the computation is performed in double precision floating
point arithmetics. A problem of this technique is that it assumes the samples
are not faulty, in the sense that they do not differ too much from each other:
an enormously different addendum would cause the average to differ as well
from the ``correct'' values. The weighted averaging technique is a partial
solution to this.

\Y\B\4\X60:Simple Average\X${}\E{}$\6
\&{static} \&{void} \\{VFA\_SimpleAverage}(\&{VotingFarm\_t} ${}{*}\\{vf},%
\39{}$\&{void} ${}{*}\\{inp}[\,],\39{}$\&{int} \\{len}${},\39{}$\&{vote\_t}
${}{*}\\{vote}){}$\1\1\2\2\6
${}\{{}$\1\6
\&{int} \|i;\6
\&{int} \|n${}\K\\{vf}\MG\|N;{}$\6
\&{double} ${}{*}\\{sum};{}$\6
\&{static} \&{char} ${}{*}\.{VFN}\K\.{"SimpleAverage"};{}$\6
\8\#\&{ifdef} \.{STATIC}\7
${}\\{sum}\K{\AND}\\{st\_VFA\_sum};{}$\6
\8\#\&{else}\6
${}\\{sum}\K\\{malloc}(\&{sizeof}(\&{double}));{}$\6
\&{if} ${}(\\{sum}\E\NULL){}$\5
${}\{{}$\1\6
${}\\{LogError}(\.{EC\_ERROR},\39\.{VFN},\39\.{"Memory\ Allocation\ E}\)%
\.{rror."});{}$\6
${}\\{VF\_error}\K\.{E\_VF\_CANT\_ALLOC};{}$\6
\&{return};\6
\4${}\}{}$\2\6
\8\#\&{endif}\C{ \PB{\.{STATIC}} }\6
\&{for} ${}({*}\\{sum}\K\T{0.0},\39\|i\K\T{0};{}$ ${}\|i<\|n;{}$ ${}\|i\PP){}$\5
${}\{{}$\1\6
${}{*}\\{sum}\MRL{+{\K}}({*}{}$(\&{double} ${}{*}){}$ \\{inp}[\|i]);\6
\4${}\}{}$\2\6
\&{if} ${}(\|n\E\T{0}){}$\5
${}\{{}$\1\6
${}\\{LogError}(\.{EC\_ERROR},\39\.{VFN},\39\.{"Inconsistency---far}\)\.{m\
cardinality\ should}\)\.{\ be\ zero."});{}$\6
${}\\{VF\_error}\K\.{E\_VF\_INVALID\_VF};{}$\6
\&{return};\6
\4${}\}{}$\2\6
${}{*}\\{sum}\MRL{{/}{\K}}\|n;{}$\6
${}\\{vote}\MG\\{outcome}\K\.{VF\_SUCCESS};{}$\6
${}\\{memcpy}(\\{vote}\MG\\{vote},\39\\{sum},\39\\{len});{}$\6
\8\#\&{ifndef} \.{STATIC}\6
\\{free}(\\{sum});\6
\8\#\&{endif}\C{ \PB{\.{STATIC}} }\6
\4${}\}{}$\2\par
\U52.\fi

\M{61}This means:
\item{$\bullet$} send a message to the server telling it
``connect me to the Agent (thread id 1)''
\item{$\bullet$} do a \PB{\\{ConnectLink}} with the Agent
\item{$\bullet$} send a set up message to the Server so that
it propagates that message to the Agent.

\Y\B\4\X61:Ask the Server to set up a connection to an Agent\X${}\E{}$\C{ yet
to be implemented }\par
\U30.\fi

\N{1}{62}Closings. This document and source code describes the actual
implementation of the Voting Farm Tool as it appears in the
EFTOS [{\sc 1}, {\sc 2}] Basic Functionality Set Library.
It has been crafted by means of the {\tt CWEB} system of
structured documentation [{\sc 3}].

\fi

\N{1}{63}Index.
Here is a list of the identifiers used, and where they appear. Underlined
entries indicate the place of definition. Error messages are also shown.

\inx

\fin
\concon
\vskip1cm

{\bf References}\vskip4mm

\item{\sc [1]} EFTOS K.U.Leuven:
{\it The EFTOS Reference Guide and Cookbook\/}.
(EFTOS Deliverable 2.4.2, March 1997)
\item{\sc [2]}
Deconinck, G., and De Florio, V., and Lauwereins, R.,
and Varvarigou, T:
EFTOS: A software framework for more dependable embedded HPC applications,
accepted for presentation at the European Conf. in
Parallel Processing (Euro-Par '97).
\item{\sc [3]} Knuth, D.E.:
{\it Literate Programming\/}
(Center for the Study of the Language and Information,
Leland Standard Junior University, 1992)
\item{\sc [4]} Carriero, N., and Gelernter, D.:
How to write parallel programs: a guide to the perplexed.
ACM Comp. Surv. {\bf 21} (1989): 323--357.
\item{\sc [5]} Carriero, N., and Gelernter, D.:
LINDA in context.
Comm. ACM {\bf 32} (1989): 444-458.

\item{\sc [6]} De Florio, V., Deconinck, G., Lauwereins, R.:
The {EFTOS} Voting Farm: a
Software Tool for Fault Masking in Message Passing Parallel Environments.
In Proc. of the 24th Euromicro Conference (Euromicro '98),
Workshop on Dependable Computing Systems, V{\"a}ster\aa{}s, Sweden, August
1998. IEEE.

\item{\sc [7]} De Florio, V., Deconinck, G., Lauwereins, R.:
Software Tool Combining Fault
Masking with
User-defined Recovery Strategies. {\it IEE Proceedings -- Software\/} {\bf
145}(6),
1998. IEE.

\item{\sc [8]} De Florio, V.:
{\it A Fault-Tolerance Linguistic Structure for Distributed Applications}.
Doctoral dissertation, Dept. of Electrical Engineering, University of Leuven,
October 2000.
ISBN 90-5682-266-7.

\item{\sc [9]} Johnson, B.W.:
{\it Design and analysis of fault-tolerant digital systems.\/}
(Addison-Wesley, New York, 1989)
\item{\sc [10]} Lorczak, P.R., and Caglayan, A.K.,
and Eckhardt, D.E.:
A Theoretical Investigation of Generalized Voters.
Proc. of the 19th Int.l
Symp. on Fault Tolerant Computing, 1989: 444--451.
%\item{$\bullet$}{\sc 11}: Truyens, M.:
%Server Network for Real-Time Parallel Applications.
%(Internal documentation, K.U.Leuven, 9 Dec. 1996)
\item{\sc [11]} Anonymous. Manual Pages of EPX 1.9.2.
(Parsytec GmbH, Aachen, 1996)
\item{\sc [12]} Anonymous. Embedded Parix Programmer's Guide.
In {\it Parsytec CC Series Hardware Documentation.\/}
(Parsytec GmbH, Aachen, 1996)

\end

%% file: cwebmac.tex
% standard macros for CWEB listings (in addition to plain.tex)
% Version 3.1 --- September 1994
\ifx\documentstyle\undefined\else \fi % LaTeX will use other macros
\xdef\fmtversion{\fmtversion+CWEB3.1}

\let\:=\. % preserve a way to get the dot accent
 % (all other accents will still work as usual)

\parskip 0pt % no stretch between paragraphs
\parindent 1em % for paragraphs and for the first line of C text

\font\ninerm=cmr9
\let\mc=\ninerm % medium caps
\def\CEE/{{\mc C\spacefactor1000}}
\def\UNIX/{{\mc U\kern-.05emNIX\spacefactor1000}}
\def\TEX/{\TeX}
\def\CPLUSPLUS/{{\mc C\PP\spacefactor1000}}
 % for backward compatibility
\def\9#1{}
\font\eightrm=cmr8
\let\sc=\eightrm % small caps (NOT a caps-and-small-caps font)
\let\mainfont=\tenrm
\let\cmntfont\tenrm
%\font\tenss=cmss10 \let\cmntfont\tenss % alternative comment font
\font\titlefont=cmr7 scaled\magstep4 % title on the contents page
 % typewriter type in title
\font\tentex=cmtex10 % TeX extended character set (used in strings)
\fontdimen7\tentex=0pt % no double space after sentences

\def\\#1{\leavevmode\hbox{\it#1\/\kern.05em}} % italic type for identifiers
\def\|#1{\leavevmode\hbox{$#1$}} % one-letter identifiers look better this way
\def\&#1{\leavevmode\hbox{\bf
  \def\_{\kern.04em\vbox{\hrule width.3em height .6pt}\kern.08em}%
  #1\/\kern.05em}} % boldface type for reserved words
\def\.#1{\leavevmode\hbox{\tentex % typewriter type for strings
  \let\\=\BS % backslash in a string
  \let\{=\LB % left brace in a string
  \let\}=\RB % right brace in a string
  \let\~=\TL % tilde in a string
  \let\ =\SP % space in a string
  \let\_=\UL % underline in a string
  \let\&=\AM % ampersand in a string
  \let\^=\CF % circumflex in a string
  #1\kern.05em}}
\def\){\discretionary{\hbox{\tentex\BS}}{}{}}
 % at sign for control text (not needed in versions >= 2.9)
\def\ATL{\par\noindent\bgroup\catcode`\_=12 \postATL} % print @l in limbo
\def\postATL#1 #2 {\bf letter \\{\uppercase{\char"#1}}
   tangles as \tentex "#2"\egroup\par}
\def\noATL#1 #2 {}
\def\noatl{\let\ATL=\noATL} % suppress output from @l
\def\ATH{\X\kern-.5em:Preprocessor definitions\X}
\let\PB=\relax % hook for program brackets |...| in TeX part or section name

\chardef\AM=`\& % ampersand character in a string
\chardef\BS=`\\ % backslash in a string
\chardef\LB=`\{ % left brace in a string
\chardef\RB=`\} % right brace in a string
\def\SP{{\tt\char`\ }} % (visible) space in a string
\chardef\TL=`\~ % tilde in a string
\chardef\UL=`\_ % underline character in a string
\chardef\CF=`\^ % circumflex character in a string

\newbox\PPbox % symbol for ++
\setbox\PPbox=\hbox{\kern.5pt\raise1pt\hbox{\sevenrm+\kern-1pt+}\kern.5pt}
\def\PP{\copy\PPbox}
\newbox\MMbox \setbox\MMbox=\hbox{\kern.5pt\raise1pt\hbox{\sevensy\char0
 \kern-1pt\char0}\kern.5pt}

\newbox\MGbox % symbol for ->
\setbox\MGbox=\hbox{\kern-2pt\lower3pt\hbox{\teni\char'176}\kern1pt}
\def\MG{\copy\MGbox}
\def\MRL#1{\mathrel{\let\K==#1}}
\let\GG=\gg

\let\NULL=\Lambda
\mathchardef\AND="2026 % bitwise and; also \& (unary operator)
 % bitwise or
 % bitwise exclusive or
 % bitwise complement
\newbox\MODbox \setbox\MODbox=\hbox{\eightrm\%}
\def\MOD{\mathbin{\copy\MODbox}}
\def\DC{\kern.1em{::}\kern.1em} % symbol for ::
 % symbol for .*
 % symbol for ->*

\newbox\bak \setbox\bak=\hbox to -1em{} % backspace one em
\newbox\bakk\setbox\bakk=\hbox to -2em{} % backspace two ems

\newcount\ind % current indentation in ems
\def\1{\global\advance\ind by1\hangindent\ind em} % indent one more notch
\def\2{\global\advance\ind by-1} % indent one less notch
\def\3#1{\hfil\penalty#10\hfilneg} % optional break within a statement
\def\4{\copy\bak} % backspace one notch
\def\5{\hfil\penalty-1\hfilneg\kern2.5em\copy\bakk\ignorespaces}% optional break
\def\6{\ifmmode\else\par % forced break
  \hangindent\ind em\noindent\kern\ind em\copy\bakk\ignorespaces\fi}
\def\7{\Y\6} % forced break and a little extra space
\def\8{\hskip-\ind em\hskip 2em} % no indentation

\newcount\gdepth % depth of current major group, plus one
\newcount\secpagedepth
\secpagedepth=3 % page breaks will occur for depths -1, 0, and 1
\newtoks\gtitle % title of current major group
\newskip\intersecskip \intersecskip=12pt minus 3pt % space between sections
\let\yskip=\smallskip
\def\?{\mathrel?}
\def\note#1#2.{\Y\noindent{\hangindent2em\baselineskip10pt\eightrm#1~#2.\par}}
\def\lapstar{\rlap{*}}
\def\stsec{\rightskip=0pt % get out of C mode (cf. \B)
  \sfcode`;=1500 \pretolerance 200 \hyphenpenalty 50 \exhyphenpenalty 50
  \noindent{\let\*=\lapstar\bf\secstar.\quad}}
\let\startsection=\stsec
\def\defin#1{\global\advance\ind by 2 \1\&{#1 } } % begin `define' or `format'
 % xref for doubly defined section name
 % xref for multiply defined section name
\def\B{\rightskip=0pt plus 100pt minus 10pt % go into C mode
  \sfcode`;=3000
  \pretolerance 10000
  \hyphenpenalty 1000 % so strings can be broken (discretionary \ is inserted)
  \exhyphenpenalty 10000
  \global\ind=2 \1\ \unskip}
\def\C#1{\5\5\quad$/\ast\,${\cmntfont #1}$\,\ast/$}
 % "// short comments" treated like "/* ordinary comments */"
%\def\C#1{\5\5\quad$\triangleright\,${\cmntfont#1}$\,\triangleleft$}
%\def\SHC#1{\5\5\quad$\diamond\,${\cmntfont#1}}
\def\D{\defin{\#define}} % macro definition
\let\E=\equiv % equivalence sign
\def\ET{ and~} % conjunction between two section numbers
\def\ETs{, and~} % conjunction between the last two of several section numbers
 % format definition
\let\G=\ge % greater than or equal sign
% \H is long Hungarian umlaut accent
\let\I=\ne % unequal sign
 % TANGLE's join operation
\let\K== % assignment operator
%\let\K=\leftarrow % "honest" alternative to standard assignment operator
% \L is Polish letter suppressed-L
\outer\def\M#1{\MN{#1}\ifon\vfil\penalty-100\vfilneg % beginning of section
  \vskip\intersecskip\startsection\ignorespaces}
\outer\def\N#1#2#3.{\gdepth=#1\gtitle={#3}\MN{#2}% beginning of starred section
  \ifon\ifnum#1<\secpagedepth \vfil\eject % force page break if depth is small
    \else\vfil\penalty-100\vfilneg\vskip\intersecskip\fi\fi
  \message{*\secno} % progress report
  \edef\next{\write\cont{\ZZ{#3}{#1}{\secno}% write to contents file
                   {\noexpand\the\pageno}}}\next % \ZZ{title}{depth}{sec}{page}
  \ifon\startsection{\bf#3.\quad}\ignorespaces}
\def\MN#1{\par % common code for \M, \N
  {\xdef\secstar{#1}\let\*=\empty\xdef\secno{#1}}% remove \* from section name
  \ifx\secno\secstar \onmaybe \else\ontrue \fi
  \mark{{{\tensy x}\secno}{\the\gdepth}{\the\gtitle}}}
% each \mark is {section reference or null}{depth plus 1}{group title}
% \O is Scandinavian letter O-with-slash
% \P is paragraph sign
\def\Q{\note{This code is cited in section}} % xref for mention of a section
\def\Qs{\note{This code is cited in sections}} % xref for mentions of a section
\let\R=\lnot % logical not
% \S is section sign
\def\T#1{\leavevmode % octal, hex or decimal constant
  \hbox{$\def\?{\kern.2em}%
    \def\$##1{\egroup_{\,\rm##1}\bgroup}% suffix to constant
    \def\_{\cdot 10^{\aftergroup}}% power of ten (via dirty trick)
    \let\~=\oct \let\^=\hex {#1}$}}
\def\U{\note{This code is used in section}} % xref for use of a section
\def\Us{\note{This code is used in sections}} % xref for uses of a section
\let\V=\lor % logical or
\let\W=\land % logical and
\def\X#1:#2\X{\ifmmode\gdef\XX{\null$\null}\else\gdef\XX{}\fi % section name
  \XX$\langle\,${#2\eightrm\kern.5em#1}$\,\rangle$\XX}
\def\Y{\par\yskip}
\let\Z=\le
\let\ZZ=\let % now you can \write the control sequence \ZZ
\let\*=*

\def\oct{\hbox{$^\circ$\kern-.1em\it\aftergroup\?\aftergroup}}% CWEB style
\def\hex{\hbox{$^{\scriptscriptstyle\#}$\tt\aftergroup}} % CWEB style
\def\vb#1{\leavevmode\hbox{\kern2pt\vrule\vtop{\vbox{\hrule
        \hbox{\strut\kern2pt\.{#1}\kern2pt}}
      \hrule}\vrule\kern2pt}} % verbatim string

\def\onmaybe{\let\ifon=\maybe} \let\maybe=\iftrue
\newif\ifon \newif\iftitle \newif\ifpagesaved

\def\lheader{\mainfont\the\pageno\eightrm\qquad\grouptitle\hfill\title\qquad
  \mainfont\topsecno} % top line on left-hand pages
\def\rheader{\mainfont\topsecno\eightrm\qquad\title\hfill\grouptitle
  \qquad\mainfont\the\pageno} % top line on right-hand pages
\def\grouptitle{\let\i=I\let\j=J\uppercase\expandafter{\expandafter
                        \takethree\topmark}}
\def\topsecno{\expandafter\takeone\topmark}
\def\takeone#1#2#3{#1}

\def\takethree#1#2#3{#3}
\def\nullsec{\eightrm\kern-2em} % the \kern-2em cancels \qquad in headers

\let\page=\pagebody \raggedbottom
\def\normaloutput#1#2#3{\ifodd\pageno\hoffset=\pageshift\fi
 \shipout\vbox{
  \vbox to\fullpageheight{
  \iftitle\global\titlefalse
  \else\hbox to\pagewidth{\vbox to10pt{}\ifodd\pageno #3\else#2\fi}\fi
  \vfill#1}} % parameter #1 is the page itself
  \global\advance\pageno by1}

\gtitle={\.{CWEB} output} % this running head is reset by starred sections
\mark{\noexpand\nullsec0{\the\gtitle}}
\def\title{\expandafter\uppercase\expandafter{\jobname}}
\def\topofcontents{\centerline{\titlefont\title}\vskip.7in
  \vfill} % this material will start the table of contents page
\def\botofcontents{\vfill
  \centerline{\covernote}} % this material will end the table of contents page
\def\covernote{}
\def\contentspagenumber{0} % default page number for table of contents
\newdimen\pagewidth \pagewidth=6.5in % the width of each page
\newdimen\pageheight \pageheight=8.7in % the height of each page
\newdimen\fullpageheight \fullpageheight=9in % page height including headlines
\newdimen\pageshift \pageshift=0in % shift righthand pages wrt lefthand ones

\def\setpage{\hsize\pagewidth\vsize\pageheight} % use after changing page size
\def\contentsfile{\jobname.toc} % file that gets table of contents info
\def\readcontents{\input \contentsfile}
\def\readindex{\input \jobname.idx}
\def\readsections{\input \jobname.scn}

\newwrite\cont
\output{\setbox0=\page % the first page is garbage
  \openout\cont=\contentsfile
       \write\cont{\catcode `\noexpand\@=11\relax}   % \makeatletter
  \global\output{\normaloutput\page\lheader\rheader}}
\setpage
\vbox to \vsize{} % the first \topmark won't be null

\def\ch{\note{The following sections were changed by the change file:}
  \let\*=\relax}
\newbox\sbox % saved box preceding the index
\newbox\lbox % lefthand column in the index
\def\inx{\par\vskip6pt plus 1fil % we are beginning the index
  \def\page{\box255 } \normalbottom
  \write\cont{} % ensure that the contents file isn't empty
       \write\cont{\catcode `\noexpand\@=12\relax}   % \makeatother
  \closeout\cont % the contents information has been fully gathered
  \output{\ifpagesaved\normaloutput{\box\sbox}\lheader\rheader\fi
    \global\setbox\sbox=\page \global\pagesavedtrue}
  \pagesavedfalse \eject % eject the page-so-far and predecessors
  \setbox\sbox\vbox{\unvbox\sbox} % take it out of its box
  \vsize=\pageheight \advance\vsize by -\ht\sbox % the remaining height
  \hsize=.5\pagewidth \advance\hsize by -10pt
    % column width for the index (20pt between cols)
  \parfillskip 0pt plus .6\hsize % try to avoid almost empty lines
  \def\lr{L} % this tells whether the left or right column is next
  \output{\if L\lr\global\setbox\lbox=\page \gdef\lr{R}
    \else\normaloutput{\vbox to\pageheight{\box\sbox\vss
        \hbox to\pagewidth{\box\lbox\hfil\page}}}\lheader\rheader
    \global\vsize\pageheight\gdef\lr{L}\global\pagesavedfalse\fi}
  \message{Index:}
  \parskip 0pt plus .5pt
  \outer\def\I##1, {\par\hangindent2em\noindent##1:\kern1em} % index entry
  \def\[##1]{$\underline{##1}$} % underlined index item
  \rm \rightskip0pt plus 2.5em \tolerance 10000 \let\*=\lapstar
  \hyphenpenalty 10000 \parindent0pt
  \readindex}
\def\fin{\par\vfill\eject % this is done when we are ending the index
  \ifpagesaved\null\vfill\eject\fi % output a null index column
  \if L\lr\else\null\vfill\eject\fi % finish the current page
  \parfillskip 0pt plus 1fil
  \def\grouptitle{NAMES OF THE SECTIONS}
  \let\topsecno=\nullsec
  \message{Section names:}
  \output={\normaloutput\page\lheader\rheader}
  \setpage
  \def\note##1##2.{\quad{\eightrm##1~##2.}}
  \def\Q{\note{Cited in section}} % crossref for mention of a section
  \def\Qs{\note{Cited in sections}} % crossref for mentions of a section
  \def\U{\note{Used in section}} % crossref for use of a section
  \def\Us{\note{Used in sections}} % crossref for uses of a section
  \def\I{\par\hangindent 2em}\let\*=*
  \readsections}
\def\con{\par\vfill\eject % finish the section names
% \ifodd\pageno\else\titletrue\null\vfill\eject\fi % for duplex printers
  \rightskip 0pt \hyphenpenalty 50 \tolerance 200
  \setpage \output={\normaloutput\page\lheader\rheader}
  \titletrue % prepare to output the table of contents
  \pageno=\contentspagenumber
  \def\grouptitle{TABLE OF CONTENTS}
  \message{Table of contents:}
  \topofcontents
  \line{\hfil Section\hbox to3em{\hss Page}}
  \let\ZZ=\contentsline
  \readcontents\relax % read the contents info
  \botofcontents \end} % print the contents page(s) and terminate
\def\concon{\par\vfill\eject % finish the section names
% \ifodd\pageno\else\titletrue\null\vfill\eject\fi % for duplex printers
  \rightskip 0pt \hyphenpenalty 50 \tolerance 200
  \setpage \output={\normaloutput\page\lheader\rheader}
  \titletrue % prepare to output the table of contents
  \pageno=\contentspagenumber
  \def\grouptitle{TABLE OF CONTENTS}
  \message{Table of contents:}
  \topofcontents
  \line{\hfil Section\hbox to3em{\hss Page}}
  \let\ZZ=\contentsline
  \readcontents\relax % read the contents info
  \botofcontents } % print the contents page(s) and terminate
\def\contentsline#1#2#3#4{\ifnum#2=0 \smallbreak\fi
    \line{\consetup{#2}#1
      \rm\leaders\hbox to .5em{.\hfil}\hfil\ #3\hbox to3em{\hss#4}}}
\def\consetup#1{\ifcase#1 \bf % depth -1 (@**)
  \or % depth 0 (@*)
  \or \hskip2em % depth 1 (@*1)
  \or \hskip4em % depth 2 (@*2)
  \or \hskip6em % depth 3 (@*3)
  \or \hskip8em % depth 4 (@*4)
  \or \hskip10em % depth 5 (@*5)
  \else \hskip12em \fi} % depth 6 or more
\def\noinx{\let\inx=\end} % no indexes or table of contents
\def\nosecs{\let\FIN=\fin \def\fin{\let\parfillskip=\end \FIN}}
    % no index of section names or table of contents
\def\nocon{\let\con=\end} % no table of contents
\def\today{\ifcase\month\or
  January\or February\or March\or April\or May\or June\or
  July\or August\or September\or October\or November\or December\fi
  \space\number\day, \number\year}
\newcount\twodigits
\def\hours{\twodigits=\time \divide\twodigits by 60 \printtwodigits
  \multiply\twodigits by-60 \advance\twodigits by\time :\printtwodigits}
\def\gobbleone1{}
\def\printtwodigits{\advance\twodigits100
  \expandafter\gobbleone\number\twodigits
  \advance\twodigits-100 }
\def\TeX{{\ifmmode\it\fi
   \leavevmode\hbox{T\kern-.1667em\lower.424ex\hbox{E}\hskip-.125em X}}}
\def\,{\relax\ifmmode\mskip\thinmuskip\else\thinspace\fi}
\def\datethis{\def\startsection{\leftline{\sc\today\ at \hours}\bigskip
  \let\startsection=\stsec\stsec}}
  % say `\datethis' in limbo, to get your listing timestamped before section 1
 % timestamps the contents page